\documentclass[11pt]{article}

\usepackage[utf8]{inputenc}
\usepackage[margin=1.3in]{geometry}
\usepackage{amsmath}
\usepackage{hyperref}
\usepackage{xcolor}
\usepackage{algorithm}
\usepackage{booktabs}
\usepackage{lscape}
\usepackage{algpseudocode}
\usepackage{etoolbox}
\usepackage{graphicx}
\usepackage{subcaption}

\usepackage[cmintegrals,cmbraces]{newtxmath}
\usepackage{ebgaramond-maths}
\usepackage[T1]{fontenc}

\makeatletter
  \DeclareSymbolFont{ntxletters}{OML}{ntxmi}{m}{it}
  \SetSymbolFont{ntxletters}{bold}{OML}{ntxmi}{b}{it}
  \re@DeclareMathSymbol{\partial}{\mathord}{ntxletters}{"40}
\makeatother

\newcommand{\orcid}[1]{\href{https://orcid.org/#1}{\includegraphics{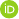}}}

\hypersetup{colorlinks,
            linkcolor={blue!80!black},
            citecolor={blue!80!black},
            urlcolor={blue!80!black}}

\title{Harnessing Implicit Cooperation: A Multi-Agent Reinforcement Learning Approach Towards Decentralized Local Energy Markets}

\author{Nelson Salazar-Peña \orcid{0009-0003-7636-4150}\textsuperscript{a}, Alejandra Tabares \orcid{0000-0002-6630-5582}\textsuperscript{b}, and Andrés González-Mancera \orcid{0000-0002-0663-2653}\textsuperscript{a,}\footnote{Corresponding author\\ \textbf{Email addresses:} na.salazar10@uniandes.edu.co (Nelson Salazar-Peña\orcid{0009-0003-7636-4150}), a.tabaresp@uniandes.edu.co
(Alejandra Tabares\orcid{0000-0002-6630-5582}), angonzal@uniandes.edu.co (Andrés González-Mancera\orcid{0000-0002-0663-2653})
}}

\date{\small{
    \textsuperscript{a} Department of Mechanical Engineering, Universidad de los Andes, 111711, Bogotá D.C., Colombia\\
    \textsuperscript{b} Department of Industrial Engineering, Universidad de los Andes, 111711, Bogotá D.C., Colombia
}}

\begin{document}

\maketitle

\section*{Abstract}

The rapid integration of distributed energy resources strains centralized management, necessitating decentralized control paradigms that ensure physical stability without compromising privacy. This paper proposes \emph{implicit cooperation}, a framework enabling decentralized agents to approximate optimal coordination in local energy markets without explicit peer-to-peer communication. We formulate the problem as a decentralized partially observable Markov decision problem that is solved through a multi-agent reinforcement learning task in which agents use stigmergic signals (key performance indicators at the system level) to infer and react to global states. Through a 3 $\times$ 3 factorial design on an IEEE 34-node topology, we evaluated three training paradigms (CTCE, CTDE, DTDE) and three algorithms (PPO, APPO, SAC). Results identify APPO-DTDE as the optimal configuration, achieving a coordination score of 91.7\% relative to the theoretical centralized benchmark (CTCE). However, a critical trade-off emerges between efficiency and stability: while the centralized benchmark maximizes allocative efficiency with a peer-to-peer trade ratio of 0.6, the fully decentralized approach (DTDE) demonstrates superior physical stability. Specifically, DTDE reduces the variance of grid balance by 31\% compared to hybrid architectures, establishing a highly predictable, import-biased load profile that simplifies grid regulation. Furthermore, topological analysis reveals emergent spatial clustering, where decentralized agents self-organize into stable trading communities to minimize congestion penalties. While SAC excelled in hybrid settings, it failed in decentralized environments due to entropy-driven instability. This research proves that stigmergic signaling provides sufficient context for complex grid coordination, offering a robust, privacy-preserving alternative to expensive centralized communication infrastructure. \newline

\footnotesize{\textbf{\textit{Keywords}} \hspace{2mm} Implicit Cooperation, Multi-Agent Reinforcement Learning, Decentralized Energy Markets, Distributed Energy Resources, Local Energy Markets}

\section{Introduction}
\label{2-sec:1-introduction}

\subsection{Motivation}
\label{2-sec:1.1-motivation}

The energy landscape is undergoing a structural transformation, driven by the decarbonization, digitalization, and decentralization \cite{charbonnier2022a}. This transition is characterized by the proliferation of distributed energy resources (DERs), such as solar photovoltaics, battery storage systems, and electric vehicles, which are transforming passive consumers into active prosumers \cite{charbonnier2022a}. While this shift promises greater grid resilience and reduced carbon emissions, it fundamentally alters traditional grid management.

The centralized control paradigm, effective for dispatching a limited number of large generators, faces intractable computational complexity and single-point-of-failure risks when attempting to coordinate thousands of geographically distributed, intermittent endpoints \cite{aoun2024}. Consequently, local energy markets (LEMs) have emerged as an operational framework for managing this complexity, enabling the decentralized trading of energy and flexibility services \cite{guerrero2019}.

However, the successful implementation of LEMs faces the challenges of achieving computational scalability to manage multiple agents, data privacy to protect agent autonomy, and the balance between supply and demand in the grid \cite{charbonnier2022a}. Traditional solutions do not satisfy all three conditions simultaneously. For example, centralized optimization fails in terms of scalability and privacy, while peer-to-peer (P2P) trading improves privacy but faces a scalability hurdle due to quadratic communication overhead \cite{charbonnier2022a}.

\subsection{Problem statement}
\label{2-sec:1.2-problem-statement}

While the necessity of decentralized coordination is well-established, current state-of-the-art approaches possess flaws regarding their deployment in real-world energy systems. The primary problem is the reliance on centralized training as the default paradigm for multi-agent learning in energy markets \cite{may2023}. While algorithms like multi-agent deep deterministic policy gradient (MADDPG) successfully demonstrate coordination in simulation, they do so by violating the privacy and decentralization constraints during the training phase. Centralized training requires a centralized critic with access to the global state including the private cost functions, battery states, and preferences of all agents to guide the learning process \cite{wilk2024}.

Conversely, decentralized training respects these privacy constraints but suffers from non-stationarity. As all agents learn simultaneously, the environment appears unpredictable to any single agent, leading to learning instability and convergence to suboptimal equilibria \cite{hady2025}. Furthermore, existing implicit coordination models often rely solely on price signals, which can induce systemic instabilities such as price volatility and load synchronization, threatening physical grid security \cite{liang2025}. 

Consequently, there is a lack of a framework that enables fully decentralized agents to learn stable, cooperative strategies for energy balance without requiring centralized training data or destabilizing price signals.

\subsection{Research gap}
\label{2-sec:1.3-research-gap}

The main challenge addressed in this paper is achieving a balance between energy supply and demand in a decentralized grid through \emph{implicit cooperation}, i.e., without relying on centralized dispatch or explicit communication. Unlike traditional centralized control, which relies on a single entity making decisions about all resources or direct negotiation, implicit cooperation requires that self-interested agents learn to work together while operating independently to achieve system-wide coherence through agents reacting to shared environmental signals \cite{moore2025}. This approach offers a potential solution by decoupling decision-making while maintaining grid balance. Subsequently, in this work we establish the theoretical basis for implicit cooperation, proposing it as a necessary coordination model for LEMs where agents must collaborate to maintain grid balance without compromising privacy or autonomy. 

This paper builds upon our previous work \cite{salazar2026}, which introduces a simulation framework for studying multi-agent interactions in LEMs and integrates modular market mechanisms with realistic physical network constraints (e.g., energy flow, congestion), to test and validate the implicit cooperation hypothesis using learning agents. This implicit cooperation challenge imposes a set of constraints that distinguish it from traditional multi-agent control problems:

\begin{itemize}
    \item \textbf{Communication constraints:} Agents must coordinate their actions without explicit, two-way communication. Coordination must emerge solely from the observation of shared environmental signals, such as agent reputation or grid congestion indicators, to preserve privacy and scalability.

    \item \textbf{Conflicting objectives:} The system is populated by self-interested agents driven to maximize their individual objective function. The challenge is to design incentive structures and information feedback loops that align individual objectives with the system-level goal of grid balance (supply-demand balance), enabling agents to learn strategies that balance both.

    \item \textbf{Non-stationarity:} The LEM environment is non-stationary because other agents are simultaneously learning and adapting. An optimal strategy for one agent depends on the strategies of others, creating a moving target problem where agents must adapt to changing opponent strategies while maintaining coordination.

    \item \textbf{Partial observability:} Agents operate with limited information about the overall state of the system. They can observe their own local state and some system-level signals, but they cannot directly observe the states or strategies of other agents. This partial observability limits agents' ability to understand the effects of their actions on the entire system.

    \item \textbf{Continuous spaces:} Unlike discrete action selection problems, energy market coordination requires continuous decisions about prices and quantities. Agents must learn policies in continuous action spaces, requiring sophisticated policy gradient methods and exploration strategies.

    \item \textbf{Uncertainty and stochasticity:} The system must account for various sources of uncertainty, including forecast errors in generation and demand, stochastic market conditions, and technical constraints. Agents must learn robust strategies that perform well under uncertainty.
\end{itemize}

\subsection{Novelty and contribution}
\label{2-sec:1.4-contribution}

To solve this challenge, we used multi-agent reinforcement learning (MARL) as our main research method, and analyzed complementary studies that provide a deeper understanding of the coordination mechanisms:

\begin{enumerate}
    \item \textbf{Training paradigms:} How do different MARL training paradigms, specifically centralized training with centralized execution (CTCE), centralized training with decentralized execution (CTDE), and decentralized training with decentralized execution (DTDE), affect the emergence and stability of implicit cooperation. This study addresses the trade-off between coordination quality and deployment feasibility, exploring whether centralized training is necessary for effective coordination or whether fully decentralized learning can achieve comparable results.

    \item \textbf{Algorithm effectiveness:} Which MARL algorithms are most effective for learning cooperative strategies in decentralized energy markets. This study explores algorithm-specific characteristics that promote or hinder coordination, providing guidance for algorithm selection in practical deployments.

    \item \textbf{Mechanism design:} How do system-level key performance indicators (KPIs) integrated into the observation space and reward function enable implicit coordination. This study delves into the mechanisms of implicit cooperation, understanding how information structure and incentive design create conditions for coordination emergence.

    \item \textbf{Scalability and robustness:} How does implicit cooperation scale with increasing numbers of agents. This study addresses practical deployment considerations, ensuring that coordination mechanisms work beyond the experimental setup.
\end{enumerate}

This research proposes that the non-stationarity of fully decentralized training (DTDE) can be resolved by engineering the observation space to include stigmergic environmental signals. Specifically, we embedded system-level KPIs (e.g., social welfare, grid congestion levels, and reputation scores) directly into the local observations of agents to create a feedback loop that functions as a surrogate for a centralized controller. This enables agents to internalize their impact on the grid and learn behaviors to balance the grid implicitly. The contributions of this work are:

\begin{itemize}
    \item We demonstrate that a decentralized implementation (DTDE), when enhanced with engineered observation spaces and reward shaping, can approximate the performance of centralized training (CTCE/CTDE), thereby validating a privacy-preserving path for real-world deployment.

    \item We introduce and validate a mechanism for using reputation scores and system-level KPIs as continuous state variables. We show how these non-price signals mitigate the synchronization risks of standard transactive energy models, allowing agents to learn stable load-balancing policies that respect physical grid constraints.

    \item We conduct a systematic evaluation of distinct training paradigm and MARL algorithm configurations in a high-fidelity simulation involving medium-scale agents with continuous action spaces. This analysis identifies the optimal algorithmic characteristics for fostering emergent cooperation in competitive energy markets.
\end{itemize}

\subsection{Paper structure}
\label{2-sec:1.5-paper-structure}

This paper provides a progression from the theoretical foundations of decentralized coordination to the computational validation of the proposed implicit cooperation model. The remainder of the paper is structured as follows: Section \ref{2-sec:2-literature-review} provides a review of the state-of-the-art in the evolution of coordination in energy systems, analyzing the limitations of explicit control in LEMs and the potential of MARL, establishing the specific gap regarding implicit cooperation and privacy. Section \ref{2-sec:3-methodology} details the methodology, covering the mathematical basis for modeling the constraints of scarce information sharing, the conceptual architecture of the implicit cooperation to incorporate system-level KPIs into the observation space and reward function to work as stigmergic signals, the theoretical analysis of the three training paradigms (CTCE, CTDE, and DTDE), and the mathematical formulations of the MARL algorithms, establishing the basis for their comparison. Section \ref{2-sec:4-experimental-setup} outlines the experimental setup, describing the framework configuration for the medium-scale agents, the grid topology, and the market rules, and the specific metrics used to assess the solution space. Section \ref{2-sec:5-results-and-discussion} presents an analysis of the experimental outcomes, comparing the efficacy of the training paradigms in fostering coordination, evaluating the robustness of the specific algorithms in continuous control tasks, and analyzing the emergent behaviors of the agents. This section provides the evidence for the trade-offs between coordination quality and deployment feasibility, validating whether fully decentralized agents can indeed approximate the performance of centralized baselines through implicit mechanisms. Finally, Section \ref{2-sec:6-conclusion} synthesizes the findings to conclude the paper, summarizing the contributions regarding the viability of implicit cooperation, offering practical guidelines for the design of privacy-preserving energy markets, and outlining future research directions for scaling these mechanisms to larger, more heterogeneous grid environments.

\section{Literature review}
\label{2-sec:2-literature-review}

\subsection{Implicit cooperation and emergent coordination}
\label{2-sec:2.2-XXX}

The evolution of power system control has been driven by the physical transformation of the grid from a centralized hierarchy to a distributed network of autonomous DERs, which has fundamentally changed the control philosophy of the electrical grid \cite{charbonnier2022a, guo2024}. Traditional centralized control, although theoretically capable of global optimization, faces intractable computational complexity (limited scalability) and creates single points of failure when applied to thousands of autonomous endpoints \cite{aoun2024, mbungu2019a}. As a result, the research landscape has shifted toward decentralized paradigms where control intelligence is distributed to the grid edge \cite{nshuti}. However, a key distinction must be made regarding how these decentralized agents coordinate: whether through explicit communication channels, involving direct negotiation and message passing, or implicit communication channels, where coordination emerges from environmental observation.

\subsubsection{A taxonomy of coordination: the limits of explicit control}

To navigate the LEM landscape, we adopt the structural taxonomy proposed by \cite{charbonnier2022a}, which classifies strategies based on the decision-making and the architecture of information flow. The fundamental distinction lies between \emph{direct control}, where a central entity issues set-points, and \emph{indirect control}, where prosumers retain autonomy \cite{charbonnier2022}.

In direct control and the paradigm of centralized bottleneck, a central controller aggregates data from all participating agents and computes global optimal set-points \cite{nshuti}. This is exemplified by the model of coordination managed by the Transmission System Operator (TSO), where the central TSO retains full responsibility for validating and dispatching DER services, treating the Distribution System Operator (DSO) merely as a data provider \cite{givisiez2020a}. While this offers high controllability and theoretical optimality, it is hindered by the curse of dimensionality \cite{charbonnier2022a}. As the number of DERs increases, the computational burden grows exponentially, and the requirement to relay vast amounts of real-time data creates significant privacy vulnerabilities and communication bottlenecks \cite{aoun2024}. Conversely, a DSO-managed model delegates validation and dispatch to the DSO, leveraging local network knowledge but potentially creating conflicts of interest regarding asset \cite{givisiez2020a}. Hybrid models attempt to bridge this by assigning pre-qualification to the DSO and dispatch to the TSO.

To address these limitations, research has pivoted to indirect control, bifurcated into \emph{mediated} and \emph{bilateral} architectures. Mediated coordination relies on a hub-and-spoke topology where a central aggregator interfaces between the grid and prosumers \cite{shafie-khah2023}. Although this facilitates value co-creation through global optimization (e.g., mixed-integer linear programming (MILP) or alternating direction method of multipliers (ADMM)), it retains a single point of failure and significant privacy vulnerabilities, as agents must disclose detailed load profiles to the coordinator \cite{charbonnier2022}. Furthermore, the computational burden of clearing complex stochastic optimization problems creates a scalability bottleneck for real-time planning \cite{anand}. Bilateral coordination, or peer-to-peer (P2P) trading, attempts to resolve this by enabling direct pairwise negotiation via technologies like blockchain \cite{guerrero2019, azim2024}. These markets often employ sophisticated mechanisms such as double auctions or game-theoretic negotiation \cite{raghoo2025}. However, this introduces a scalability wall since the communication overhead in a mesh network grows quadratically with the number of participants \cite{guo2024}. Approaches relying on explicit preference matching often incur prohibitive negotiation latency, making them unsuitable for the granular, intraday planning required by thousands of assets \cite{charbonnier2022a}.

\subsubsection{Price-only mechanisms for implicit coordination}

The architectural choice of a control system defines the operational capabilities and failure modes of a LEM. The literature delineates a spectrum ranging from fully centralized to fully decentralized  (see Table \ref{tab:control-comparison}), each presenting distinct trade-offs between system-level optimality and agent-level autonomy.

At one end of the spectrum lies the centralized control architecture, typically orchestrated by a central controller \cite{nshuti}. In this paradigm, the central controller aggregates data from all participating agents (e.g., generators, loads, and storage units) and computes global optimal set-points to minimize operational costs or maintain stability \cite{nshuti}. While this approach theoretically ensures global optimality and high controllability, it is increasingly viewed as computationally intractable for large-scale systems involving high DER penetration \cite{aoun2024}. Furthermore, centralized architectures introduce a single point of failure. The compromise or malfunction of the central controller can cascade into system-wide collapse, and the communication infrastructure required to relay vast amounts of real-time data creates significant bottlenecks and privacy vulnerabilities \cite{aoun2024}.

Conversely, decentralized control paradigms distribute decision-making authority to individual agents or local controllers \cite{nshuti}. This architecture aligns with the physically distributed nature of DERs, enabling agents to make decisions based on local measurements (e.g., voltage, frequency) or local objectives \cite{kantamneni2015}. The primary advantages of decentralization are enhanced resilience as the system lacks a single point of failure, and plug-and-play scalability, which allows for the integration of new assets without redesigning the central control logic \cite{aoun2024, kantamneni2015}. However, purely decentralized decisions, if uncoordinated, can lead to suboptimal global outcomes or localized imbalances \cite{alazemi2022a}. P2P control represents a further evolution of this autonomy, where agents act as peers without hierarchical relationships, forming the technical basis for direct energy trading \cite{nshuti, ZHENG2024133}.

Systematic reviews indicate a trend toward decentralized coordination schemes in this domain \cite{alazemi2022a}. Centralized optimization is increasingly viewed as practically infeasible due to data privacy concerns and computational load. Decentralized approaches, often employing game-theoretic or distributed optimization, are favored because they significantly reduce complexity and operational costs, even if they yield theoretically sub-optimal resource allocations compared to a perfect centralized benchmark \cite{alazemi2022a}. This real-world structural trend reinforces the need for scalable, low-communication coordination methods in LEMs.

To manage the decentralized actions of prosumers within LEMs, coordination strategies have shifted from direct physical commands to market-based mechanisms that utilize economic signals. For instance, demand response represents the foundational layer of this shift, incentivizing consumers to alter consumption patterns in response to price signals or reliability needs \cite{mbungu2019a}. While effective for peak shaving, incentive-based demand response faces technical challenges in establishing accurate counterfactual baselines for consumption, which can be strategically exploited by participants \cite{papavasiliou2010, satchidanandan2023}.

Transactive energy extends these principles by creating a dynamic equilibrium between supply and demand across the entire infrastructure, using profit as the key operational parameter \cite{khaskheli2024}. Transactive energy empowers prosumers to actively buy and sell energy, improving efficiency and integrating intermittent renewables \cite{huang2021a, zhou2023a}.

P2P energy trading is the most decentralized manifestation of transactive energy, facilitating direct exchange between agents \cite{guerrero2019}. These markets often employ sophisticated mechanisms such as double auctions, game-theoretic negotiation, or optimization-based clearing \cite{raghoo2025}. However, these approaches predominantly rely on explicit coordination: they require structured, binding communication protocols (e.g., submitting formal bids and asks) and complex market clearing processes \cite{wang2022a}. This reliance on explicit negotiation creates significant communication overhead, computational latency during clearing, and potential privacy concerns regarding the disclosure of prosumer preferences \cite{guerrero2019, anderson2011}.

\begin{table}[h]
\centering
\caption{Comparative analysis of control paradigms for distributed energy resources (DERs).}
\label{tab:control-comparison}
\resizebox{\textwidth}{!}{%
\begin{tabular}{@{}lllllll@{}}
\toprule
\textbf{Paradigm} & \textbf{Controller} & \textbf{Coordination Signal} & \textbf{Key Advantage} & \textbf{Key Disadvantage} & \textbf{Scalability} & \textbf{Resilience} \\ \midrule
Centralized control & Single central controller & Direct command and control signals & Global optimization, high controllability & Single point of failure, high computational burden & Low & Low \\

Decentralized control & Individual agents & Local measurements (voltage, frequency) & No single point of failure, plug-and-play capability & Suboptimal global outcomes, system imbalance & High & High \\

Transactive energy & Individual agents & Economic signals (prices, bids, asks) & Fosters prosumer autonomy, economic efficiency & Communication infrastructure, potential for market power abuse & High & High \\ \bottomrule
\end{tabular}%
}
\end{table}

\subsubsection{Theoretical foundations: from explicit negotiation to stigmergy}

The challenges of price-only signals necessitates a shift toward more robust forms of implicit coordination, grounded in the concept of stigmergy. Stigmergy describes a mechanism of indirect coordination where an action taken by an agent leaves a trace in the environment, which subsequently stimulates the performance of a next action by the same or a different agent \cite{frey2015, schubotz2022, denicola2020}.

In the context of a LEM, the market platform itself functions as the stigmergic environment \cite{denicola2020}. Agents do not need to communicate directly (distributed topology) or report to a central controller (centralized topology). Instead, they interact with the shared environment. An agent's action (such as injecting or withdrawing power) modifies shared environmental variables. These modifications serve as environmental markers \cite{schubotz2022, deng2019}. Other agents perceive these changes and adapt their policies accordingly \cite{ziras2019}. This process allows for system-wide order to emerge from local rules, effectively decoupling the complexity of the system from the complexity of the individual agents \cite{sauter2005, roca2016}.

The objective of this interaction is emergent coordination: the realization of a global property (such as supply-demand balance) that is not explicitly programmed into any single agent but arises from their collective interactions \cite{roca2016}. However, emergence is non-deterministic and not inherently beneficial; without careful design, the aggregation of self-interested local rules can lead to grid instability \cite{binder2019}. Therefore, the research challenge shifts towards engineering emergence via quantitative, marker-based stigmergy \cite{schubotz2022, debe2021}. By embedding system-level KPIs into the local observation space of each agent, the environment effectively guides the learning process. Agents trained via MARL can learn to interpret these environmental markers as predictors of future rewards, leading to the emergence of cooperative policies even in the absence of explicit cooperation protocols \cite{verborgh2021}.

\subsubsection{Emergent coordination and the challenge of unintended behaviors}

The objective of implicit cooperation is emergent coordination: the realization of a coherent global property such as supply-demand balance that is not explicitly programmed into any single agent but arises organically from their collective interactions. This principle suggests that for large-scale systems, it is neither feasible nor necessary to explicitly orchestrate every decision for every scenario. Instead, desirable system-wide behaviors can be induced by designing the right set of local rules and information structures \cite{roca2016}. However, achieving beneficial emergence requires navigating a complex landscape of coordination architectures, where the limitations of explicit control highlight the necessity for implicit mechanisms.

In mediated coordination architectures, where a central entity manages the system, the computational burden becomes prohibitive. \cite{lezama2019} illustrates this through a stochastic optimization framework that bridges wholesale and local markets; while mathematically robust, the need to manage uncertainty via Monte Carlo simulations results in an explosion of scenarios, rendering the central optimization problem computationally intractable for real-time, large-scale applications. Similarly, \cite{tsaousoglou2022} highlight the latency issues inherent in Lagrangian decomposition methods. While these iterative pricing mechanisms theoretically achieve convergence, the communication delays required for agents to compute and upload responses, coupled with a privacy challenge where iterative exchanges eventually reveal private utility functions, make them ill-suited for the granular responsiveness required in modern grids.

Decentralizing control via bilateral coordination (P2P) resolves the single point of failure but introduces severe communication overhead. \cite{guerrero2019} proposes a network-constrained P2P scheme that utilizes sensitivity coefficients (voltage sensitivity coefficients and power transfer distribution factors) to validate trades against grid physics. While this successfully integrates physical constraints, the requirement to validate every bilateral transaction creates a new computational bottleneck. Furthermore, the negotiation process itself induces significant latency. \cite{paudel2019} demonstrates this in a game-theoretic model where buyers and sellers engage in nested evolutionary and Stackelberg games. The time required for these iterative algorithms to converge to a stable strategy makes them impractical for high-frequency, intraday planning. Even when mechanisms evolve to include heterogeneous preferences, as seen in \cite{talari2022}, where agents trade based on attributes like reputation and location, the reliance on explicit matching algorithms retains a heavy communication burden that mirrors the negotiation complexity.

Consequently, the research challenge shifts from explicit orchestration to engineered implicit coordination. Here, agents do not negotiate, they merely observe and react. However, the literature cautions that undefined emergence is non-deterministic and not inherently beneficial. Without careful design, the aggregation of self-interested local rules can lead to grid instability. If thousands of agents react independently to a simple broadcast signal (like a low price) without coordination logic, they may synchronize their actions creating a new emergent peak that destabilizes the network \cite{roca2016, binder2019}.

To mitigate this, the information structure of the environment must be designed so that locally optimal actions align with global stability. \cite{charbonnier2022} provides a breakthrough in this domain using MARL. Their framework demonstrates that by training agents with marginal rewards (signals that quantify an individual's specific contribution to the global state) agents can learn to desynchronize their behaviors. Through Q-learning, agents learn to interpret environmental markers not just as immediate incentives, but as predictors of collective system states, effectively learning to wait or curtail consumption voluntarily when the grid is congested \cite{verborgh2021}.

The frontier of this field involves expanding the observation space beyond simple price signals to include the rich, non-price attributes identified in explicit models. The physical sensitivities identified by \cite{guerrero2019} and the reputation and preference metrics utilized by \cite{talari2022} must be transformed from validation tools into passive environmental signals. By embedding these system-level objectives into the local observation space, the environment effectively guides the learning process, allowing complex cooperative policies to emerge without the scalability constraints of direct communication.

\subsubsection{KPI-based coordination and the role of reputation in energy markets}

A primary vehicle for implementing quantitative stigmergy in LEMs is the use of KPIs and reputation metrics. The shift from explicit negotiation to implicit coordination requires a specific mechanism to serve as the environmental signal. While price signals have proven unstable, reputation serves as a robust parameter that encodes historical performance into an environmental marker \cite{schubotz2022}. By turning trustworthiness into a tangible, data-based measure, reputation systems allow agents to coordinate implicitly, aligning with system goals not through direct commands but through the drive for higher standing and the associated economic advantages.

In the context of multi-agent systems, it is necessary to distinguish reputation from trust. Trust is often defined as a subjective, private belief formed through direct pairwise interaction \cite{ramchurn2004}. While early models relied on subjective recommendations (e.g., agents rating each other), this approach is vulnerable to bias and collusion \cite{wang2003}. Conversely, reputation is a collective, public measure of an agent's standing derived from observable historical performance \cite{pujol2002}. For LEMs, a robust paradigm relies on quantitative, marker-based coordination where reputation is a verifiable fact calculated from objective KPIs \cite{schubotz2022}. This approach aligns with research incorporating social network topology, where factors like an agent's relative electrical distance contribute to its reputation independent of direct feedback \cite{pujol2002}.

To function as an effective coordination signal, reputation must be quantified through multi-faceted metrics that capture distinct dimensions of agent behavior, spanning physical reliability to cyber-security. For instance, the commitment score measures the reliability of an agent by tracking the deviation between cleared market schedules and actual power delivery \cite{nair2025}. Calculated as a moving average of these deviations, it provides a direct, historical record of physical compliance, determining whether a prosumer will actually deliver the flexibility they agreed. Furthermore, the trustability score safeguards against compromised nodes by utilizing anomaly detection on cyber-physical data (such as communication packet metadata and power data) to assess the integrity of an agent \cite{nair2025}.

Reputation can also be directly linked to grid constraints. For instance, a voltage impact score penalizes prosumers who contribute to local voltage rise while rewarding those who mitigate it, effectively internalizing grid physics into social standing \cite{yapa2024}. Similarly, system KPIs such as grid congestion levels serve as observable signals \cite{yapa2024}. Literature proposes holistic metrics like the resilience score \cite{nair2025}, which aggregates the commitment and trustability scores, or the R360 score, which integrates quality of service, defense capabilities, and resource availability \cite{sang2022}.

Once quantified, these scores transition from passive indicators to active control variables that close the loop in implicit coordination. In decentralized P2P markets, they act as a filter for partner selection, where agents prefer transacting with high-reputation peers to minimize risk \cite{yapa2024}. This effectively marginalizes unreliable agents and prevents a race to the bottom on price. Advanced models even propose reputation-adjusted pricing, where high-reputation agents receive preferential rates, creating a direct economic incentive for grid-friendly behavior \cite{yapa2023}.

More critically for MARL applications, these metrics function as state variables that resolve the non-stationarity of the environment. In a fully decentralized setting, other agents appear as part of the environment, making it unpredictable. By making reputation scores and system KPIs visible in the observation space, agents can learn to correlate these high-level signals with the probability of successful trades or grid constraint violations, thereby making the environment more predictable.

However, the reliance on reputation introduces new vulnerabilities. The robustness of the system against strategic manipulation remains a primary research challenge. The literature identifies risks such as playbook attacks, where agents build a high reputation through numerous low-value transactions only to exploit this trust for a massive defection on a high-value trade \cite{josang, raza2024}. Simple moving-average metrics like the commitment score are vulnerable to this. Therefore, robust metric design must weight actions by their impact and criticality, not just their frequency.

Furthermore, agents may engage in collusion to artificially inflate scores, or the system may suffer from emergent centralization, where a few agents dominate the market because high reputations lead to more interaction opportunities \cite{josang}. A resilient design must balance efficiency with diversity, potentially by incentivizing interaction with new or mid-tier agents to foster system-wide resilience.

Current analysis confirms that while reputation provides the necessary trust layer for implicit coordination, naive implementations are insufficient. The gap in the state of the art lies in the lack of implicit coordination models that specifically address the real-time supply-demand balancing problem through stigmergic mechanisms. Existing approaches largely restrict reputation to a filtering mechanism for explicit trading or fault detection, rather than using it as the primary environmental signal to drive cooperative load-balancing behaviors \cite{khaskheli2024}. Furthermore, there is a lack of systematic comparison regarding how different training paradigms (centralized versus decentralized) affect the emergence of this implicit cooperation when agents are driven by these shared KPI signals.

\subsection{MARL in energy systems}
\label{2-sec:2.2-marl-energy-systems}

The structural transformation driven by the decentralization of modern power systems has rendered traditional centralized optimization paradigms insufficient, primarily due to their inability to manage the stochasticity and scale of DERs effectively \cite{sanayha2022, kerena}. As renewable energy penetration increases, the grid is evolving from a passive distribution network into an active ecosystem of prosumers, microgrids, and aggregators, necessitating robust LEMs that facilitate P2P and community-based trading \cite{kerena}.

In this context, MARL has emerged as a critical computational framework. Unlike centralized optimization methods such as convex optimization or dynamic programming, which struggle with the non-linear dynamics and computational burden of coordinating thousands of heterogeneous agents, MARL enables decentralized agents to learn optimal policies through interaction with the environment \cite{kerena}. However, the application of MARL in LEMs introduces complex challenges, particularly regarding coordination under uncertainty. The non-stationary nature of the environment, where the optimal strategy of one agent depends on the changing strategies of others, requires sophisticated learning paradigms beyond simple independent learning \cite{kerena}. Furthermore, the physical constraints of the power grid and the privacy requirements of market participants impose severe limitations on information sharing, creating a need for implicit cooperation models where coordination is achieved without direct communication \cite{lib}.

\subsubsection{From independent learning to centralized training}

The application of RL to energy systems has evolved in parallel with the complexity of the grid itself. Early research predominantly utilized independent learning approaches, such as independent Q-learning \cite{charbonnier2022a}. In this paradigm, each agent (e.g., a home energy management system or a battery controller) views other agents merely as part of the environment. While independent learning is computationally efficient and scales linearly with the number of agents, it fundamentally violates the Markov property required for convergence: as agents simultaneously update their policies to maximize local rewards, the environment becomes non-stationary from the perspective of any single agent \cite{kerena}.

This non-stationarity leads to undesired behaviors where agents chase a moving target, leading to policy oscillations. For instance, if multiple prosumers learn to discharge batteries during a price spike, their collective action may eliminate the arbitrage opportunity, invalidating the policy they just learned \cite{kerena}. Furthermore, independent learners in partially observable environments often converge to suboptimal Nash equilibria that fail to achieve global Pareto optimality, potentially causing grid instability issues such as voltage oscillations \cite{charbonnier2022a}. Moreover, in cooperative settings, determining which agent's action contributed to a global reward (e.g., grid balance) is difficult without a mechanism to decouple individual contributions \cite{kerena}.

To mitigate these instabilities without reverting to intractable centralized control, the research community has come together around the CTDE paradigm \cite{amato2025}. CTDE acknowledges the dual nature of the problem: training can occur with access to global state information, while execution must rely solely on local observations. This information is used to train a centralized critic, which then guides the updates of decentralized actors that execute using only local observations \cite{hady2025}.

Prominent algorithms like MADDPG use a centralized critic to estimate Q-values based on the joint state and action space, stabilizing the learning process for decentralized actors \cite{charbonnier2022a, lowe2020}. MADDPG employs a centralized critic that takes the joint action and state space as input to estimate Q-values, while actors remain decentralized. Fundamentally, a centralized critic is a training-only neural nerwork capable of accessing the full global state and joint actions, and Q-values quantify the expected long-term return of taking a specific action in a given state. Studies have demonstrated MADDPG's effectiveness in continuous action spaces (crucial for energy bidding), showing it can flatten community demand curves and reduce costs more effectively than independent learners \cite{wilk2024}.

While MADDPG is off-policy (i.e., it updates its policy using stored data collected by older versions of the agents or a completely different exploration strategy), multi-agent proximal policy optimization (MAPPO) applies the on-policy PPO algorithm to multi-agent settings, requiring that an agent updates its current policy based on the actions it takes while following that same policy \cite{schulman2017a}. This difference impacts convergence. Recent comparative studies in industrial manufacturing and grid control suggest MAPPO often exhibits superior stability and hyperparameter robustness compared to MADDPG, particularly in high-dimensional environments where off-policy learning can suffer from distributional shift \cite{ahilan2019}.

Similarly, value-decomposition methods like QMIX, which addresses the multi-agent credit assignment problem by breaking down a global team reward into individual agent contributions, have become standard for discrete control tasks. They function by decomposing the global team reward into individual agent utilities, ensuring that a maximization of local values leads to a maximization of the joint value \cite{wilk2024}.

However, the standard CTDE framework faces challenges in LEMs regarding data privacy and scarce information sharing. The requirement for a centralized critic implies that during training, all agents must share their private states (e.g., battery levels, consumption patterns, cost functions) with a central learner \cite{khaskheli2024}. This contradicts the privacy requirements of many deregulated LEMs and the operational reality where agents may be competitive or owned by distinct entities, necessitating the exploration of implicit cooperation models where even the training phase minimizes global information exchange \cite{charbonnier2022a}.

\subsubsection{Training paradigms for MARL}

While CTDE is the preferred choice in the literature review, a critical analysis reveals significant limitations when applied to privacy-sensitive LEMs. The choice of training paradigm dictates the information flow between agents and constrains the type of coordination (explicit versus implicit) that can be achieved. A systematic comparison of the three primary paradigms reveals the specific gap this research addresses:

\begin{itemize}
    \item \textbf{CTCE:} This paradigm treats the multi-agent problem as a single-agent problem with a joint action space. A central controller (i.e., a single neural network) receives the full state containing all private information of all agents and optimizes the joint policy (i.e., the action of each agent) \cite{amato2025}. While this serves as a theoretical upper bound for performance, its challenge for LEMs is due to the scalability bottleneck (exponential growth as the number of agents increase) and privacy violation since prosumers are increasingly unwilling to share detailed usage data and cost functions with a central authority \cite{amato2025}.

    \item \textbf{CTDE:} While solving the execution-time communication problem, CTDE assumes rich information sharing during training. Agents possess individual actor neural networks that execute locally based on partial observations. However, during training, a centralized critic neural network is utilized \cite{hady2025}. The requirement for a centralized critic implies that during the learning phase, all agents must share their private observations (e.g., battery state, consumption preferences, internal cost functions) with a central learner \cite{hady2025}. However, the CTDE facilitates implicit cooperation by allowing the critic to train the actor how their local action contributes to the overall reward. The actor learns to approximate the global value function using only local data. However, for a LEM where agents may be competitive, sharing private observations for a centralized critic is often prohibitive \cite{khaskheli2024}.

    \item \textbf{DTDE:} This paradigm aligns best with the constraints of effective training without global view assumptions \cite{amato2025}. In DTDE, agents update policies based solely on their own trajectories (observations, actions, and rewards). While this respects the decentralized partially observable Markov decision process (Dec-POMDP) constraints (i.e., limiting agents to local partial observations without access to the true global state), it reintroduces the challenges of non-stationarity and coordination without communication \cite{ahilan2019}. The frontier of research lies here: developing mechanisms that allow DTDE agents to achieve coordination comparable to CTDE without the centralized information requirements \cite{charbonnier2022a}. Recent work, such as \cite{wilk2024}, attempts to bridge CTDE and DTDE by sharing only strategic information rather than raw data, optimizing decision-making while enhancing privacy. This represents a move towards federated or privacy-preserving MARL \cite{wen2021a}, but rigorous comparisons in continuous LEMs remain limited.
\end{itemize}

The Dec-POMDP framework is uniquely suited for LEMs because it explicitly accounts for the partial observability caused by privacy constraints and distributed geography \cite{kerena}. An agent knows its own load and battery state but cannot observe the internal state of other agents \cite{lauri2020}. The global state (network voltage, total demand) is a result of joint actions but is not fully visible to any single entity \cite{amato2013}. Solving Dec-POMDPs necessitates the use of approximate solutions via MARL rather than exact dynamic programming \cite{amato2013}. Recent works have explicitly modeled P2P trading as a Dec-POMDP, utilizing RL to approximate optimal policies in continuous spaces where traditional heuristics fail \cite{lou2025}.

\subsubsection{Mechanisms for implicit cooperation in MARL}

In scarce information scenarios where explicit communication is restricted, MARL agents must coordinate via implicit cooperation. This relies on agents learning to interpret environmental feedbacks as coordination signals. The physics of the grid and the economics of the market act as natural stigmergic media. Deviations in grid frequency or voltage act as implicit broadcast signals. A drop in frequency signals all agents to increase generation or shed load. MARL agents can learn to interpret these environmental feedbacks to coordinate corrective actions (e.g., voltage restoration) without explicit data exchange \cite{wu2021}. Similarly, the clearing price in a LEM aggregates the preferences and scarcity of all agents. A high clearing price implicitly communicates a state of high demand or low supply. Research demonstrates that MARL agents can learn to use price volatility as a coordination signal, for example, inferring that a sharp price rise indicates a neighbor is discharging, prompting the agent to hold its own capacity \cite{zhu2022}. This mechanism allows agents to achieve load balancing simply by reacting to the aggregate price signal \cite{charbonnier2022a}.

Recent MARL research studies how self-interested agents in mixed-motive games like LEMs can learn reciprocal altruism. Agents may learn that unrestrained greed leads to market collapse or grid instability, and thus long-term penalties. Through repeated interaction, agents can discover sustainable strategies where they voluntarily curtail consumption during peaks to lower the clearing price for everyone \cite{wilk2024}. To foster these behaviors, reward shaping is often employed, modifying the agent's reward function to include not just individual profit, but also a marginal contribution to the system or a global stability term \cite{stein2024}. This technique aligns individual greed with global welfare, enabling cooperative behaviors such as peak shaving to emerge from self-interested learning \cite{phan, mannion2018}.

In the absence of direct communication, agents must construct internal models of other agents. Techniques like deep implicit coordination graphs proposed in \cite{li2021a} allow agents to infer the hidden states or future intentions of peers based on their observable actions \cite{wang2025}. For instance, an agent might infer that a competitor's aggressive bidding implies a high state-of-charge, subsequently adjusting its own strategy to avoid a price competition. This inference allows agents to form a consensus about the system state based on local observations \cite{lib}.

\subsubsection{Review of existing MARL applications in LEMs}

The application of MARL to energy systems has expanded to address specific market structures and physical constraints. Research in P2P markets has utilized consensus MARL to enable agents to develop trading strategies that converge to dynamic balance and voltage stability without central dispatch \cite{feng}. Other works utilize local strategy-driven MADDPG to harmonize individual building needs with community goals, effectively reducing energy costs and flattening demand curves through shared strategic information \cite{wilk2024}. For microgrids, multi-agent twin delayed DDPG (MATD3) has been used to optimize the bidding behavior of energy storage units, offering better stability than standard MADDPG in market modeling \cite{wolgast2024}. Hierarchical approaches have also been proposed, where an upper-level agent (aggregator) learns to coordinate lower-level agents (households), although this often leans back towards centralized control \cite{zhang2025}.

The literature distinguishes between agent scales, which impacts the learning formulation. While much research is focused on residential households, the medium-scale agent represents a distinct and critical class of actors in LEMs. A medium-scale agent refer to entities larger than a single residential household but smaller than utility-scale power plants, such as commercial buildings and industrial microgrids \cite{lu2024, qiua, guo2025, feng2024a}. Unlike small households which are price-takers, medium-scale agents often possess sufficient capacity to be price-makers \cite{lu2024}. This introduces market power issues where the agent's action directly impacts the state transition of the market price, reinforcing the non-stationary nature of the environment \cite{li2024}. For instance, an aggregator controlling a large fleet of electric vehicles can significantly alter the clearing price by withholding or flooding the market. MARL is particularly valuable here for learning strategic bidding behaviors that account for this price impact, optimizing the trade-off between volume and price influence \cite{eusebio2016}.

For medium-scale, control inherently involves continuous variables (power quantities and prices) where an agent must decide how much energy to trade and at what price. Early approaches used deep Q-network (DQN) with discretized actions but results in the curse of dimensionality and loss of precision, leading to suboptimal market clearing \cite{kerena}. In the industrial sector, MARL has been applied to optimize energy consumption in energy-intensive settings using POMDP frameworks \cite{bashyal2025}. For commercial buildings, MARL agents control HVAC and storage to minimize costs under dynamic pricing. Notably, research comparing MAPPO and MADDPG in these settings suggest that while MADDPG excels in continuous action manipulation, MAPPO often provides more robust convergence in complex, multi-modal environments \cite{gao2025}. Furthermore, medium-scale agents operate critical infrastructure with strict physical limits. Safe (or constrained) MARL is a growing sub-field where agents learn policies that maximize reward while satisfying safety constraints. Techniques such as Safe-MADDPG incorporate barrier functions or Lagrangian multipliers into the loss function to enforce constraints (e.g., voltage limits) during the learning process \cite{kallstrom2023}. This is relevant for coordination without explicit communication, ensuring that implicit cooperation does not violate grid safety limits \cite{mylonas2025}.

Recent research favors policy gradient methods like MADDPG and soft actor-critic (SAC), which can output continuous actions directly \cite{charbonnier2022a}. Recent literature has provided comparative analyses of algorithms for continuous energy problems. SAC is increasingly favored for its maximum entropy framework, which encourages exploration, a critical feature for non-stationary LEMs where agents must constantly adapt to the changing policies of competitors \cite{zhu2022a}. SAC has been shown to outperform DDPG in robustness and sample efficiency for microgrid energy management \cite{sharma2025}.

Despite the efforts of existing research, there is a lack of systematic comparison of training paradigms (CTCE, CTDE, DTDE) specifically evaluated on their ability to foster coordination under scarce information constraints \cite{ahilan2019}. Most studies default to CTDE (e.g., MADDPG) without questioning the privacy implications of sharing observations during training \cite{kerena}. The trade-off between the optimality of CTDE and the privacy/scalability of DTDE remains under-explored \cite{ahilan2019}. Furthermore, the majority of literature focuses either on micro-scale households or macro-scale TSO operations. The coordination of medium-scale agents via implicit cooperation is under-researched \cite{charbonnier2022a}. These agents possess market power, meaning their actions directly impact price formation, reinforcing the non-stationary nature of the environment and requiring specialized MARL formulations (e.g., Safe-MADDPG or constrained SAC) rather than generic P2P algorithms \cite{li2024, azizanruhi2018}. Moreover, while Dec-POMDPs are the theoretical standard for partial observability \cite{kerena}, applied research often simplifies the problem to discrete action spaces or assumes full observability. There is a distinct need for robust MARL algorithms that act within a rigorous continuous Dec-POMDP formulation for intraday markets, specifically addressing coordination without direct communication through implicit mechanisms rather than explicit message passing \cite{lauri2020}.

\subsection{Training paradigms: theory and practice}
\label{2-sec:2.3-training-paradigms}

The transition from centralized dispatch models toward decentralized grid operation and control introduces stochasticity and complexity that traditional optimization techniques struggle to manage due to computational burdens and privacy constraints \cite{charbonnier2022a}. Consequently, MARL has emerged as a framework for enabling autonomous agents to learn optimal policies in these complex environments. However, the deployment of MARL in LEMs is constrained by the architecture chosen for information sharing and decision authority. This architectural choice defines the system’s ability to balance the coordination of DERs for energy management considering the complexities of computational scalability, performance optimality, and data privacy \cite{charbonnier2025, qiu2023a}. The literature outlines a range of training paradigms (CTCE, CTDE, and DTDE) each with different theoretical features and practical trade-offs concerning this challenges (see Table \ref{tab:training-taxonomy}).

\subsubsection{CTCE: the theoretical benchmark for optimality}

CTCE represents the classical top-down approach. In this paradigm, a single, omniscient controller possesses a complete global view, gathering all private observations, rewards, and state information from every agent to solve a global optimization problem \cite{charbonnier2025}. In the context of LEMs, CTCE serves as the theoretical performance upper bound benchmark \cite{charbonnier2025}. It is frequently implemented via MILP or single-agent RL to quantify the loss of efficiency inherent in decentralized systems compared to a central planner with perfect information \cite{charbonnier2025, velvis}. While CTCE is theoretically ideal for calculating the social welfare upper bound, the literature is unanimous in classifying it as non-viable for real-world deployment due to three disqualifying failures \cite{qi}:

\begin{itemize}
    \item \textbf{Computational intractability:} The joint state-action space grows exponentially with the number of agents, known as the curse of dimensionality. This renders optimization computationally intractable for all but small problems \cite{charbonnier2022a}.

    \item \textbf{Privacy violation:} CTCE requires prosumers to surrender granular, high-resolution data such as internal consumption patterns, battery states-of-charge, and economic utility functions to a central monopoly. This is an undesirable requirement in deregulated, privacy-focused markets \cite{dao2024}.

    \item \textbf{Latency and infrastructure constraints:} The entire system's reliability hinges on the continuous, flawless operation of a single central controller \cite{kerena}. This architecture is brittle and vulnerable to communication failures. Research indicates that fault protection in microgrids often requires reaction times of 3–5 ms; centralized controllers relying on standard telemetry frequently fail to meet these stringent limits, creating a risk for critical energy infrastructure \cite{muyizere2022}.
\end{itemize}

\subsubsection{CTDE: the hybrid standard for stability}

To address the limitations of centralized execution, the research community has come together around the CTDE as the current academic standard \cite{ikeda2022}. Exemplified by algorithms like MADDPG and MATD3, this paradigm distinguishes between a centralized training phase and a decentralized execution phase \cite{charbonnier2025}.

CTDE algorithms solve the non-stationarity inherent in multi-agent learning by employing a centralized critic during training \cite{wilk2024}. This critic receives the joint observations and actions of all agents, effectively making the environment stationary from its perspective. By observing the full global state, the critic can explicitly solve the credit assignment problem, determining exactly how agent's action contributed to the global reward amidst the noise of other agents' behaviors \cite{charbonnier2025}. This allows decentralized actors to learn coordinated policies, as in \cite{charbonnier2022a}, where agents implicitly desynchronize their load profiles to avoid price competition in storage markets.

However, the centralized critic’s neural network architecture typically has a fixed input layer dimension that is a function of the number of agents. If a new prosumer joins the LEM, or an existing agent leaves, the input vector size changes, rendering the trained model invalid. The system cannot adapt. Thus, it must be taken offline, the architecture redesigned, and the entire system retrained from scratch \cite{charbonnier2025}. Furthermore, the training assumes the existence of a high-fidelity simulator where global states (including private prosumer data) are available. In privacy-sensitive residential scenarios, constructing such a centralized training environment is often as unfeasible as centralized execution \cite{charbonnier2025, yang2025}. Thus, while CTDE solves the execution-time and single point of failure, it fails to provide the plug-and-play scalability and privacy required for dynamic energy markets.

\subsubsection{DTDE: the frontier of decentralization}

The disqualification of centralized paradigms forces the conclusion that DTDE (also known as independent learning) is the viable architecture for scalable, private, and robust LEMs \cite{hu2024}. In DTDE, agents learn and act based solely on local data, ensuring linear scalability and data sovereignty \cite{charbonnier2025}.

Despite its architectural suitability, naive DTDE fails in practice due to the non-stationarity problem. From the perspective of any single agent, the environment is incoherent because all other agents are simultaneously updating their policies \cite{charbonnier2025}. This violation of the Markov property leads to the credit assignment failure where an agent cannot distinguish whether a reward (e.g., a profitable trade) was due to its own intelligent bid or simply random noise resulting from the simultaneous actions of others \cite{charbonnier2025}, and also to the avalanche effect where independent agents optimizing for similar tariffs inevitably synchronize their behaviors. For example, in residential energy management, independent learners often identify the same low-price interval and synchronize their battery charging. This collective action creates a new maximum load, impacting arbitrage opportunities and violating network restrictions \cite{charbonnier2022a, rothkopf2007}.

The failure of DTDE stems from a partially observable environment where agents are blind to the systemic consequences of their actions. To safeguard DTDE's scalability, agents must coordinate. However, explicit communication is disqualified as it re-introduces high bandwidth costs and privacy risks \cite{qiu2021a}.

The proposed solution lies in implicit coordination, where MARL agents can coordinate through observation augmentation \cite{charbonnier2025}. This stigmergy allows for a common frame of reference, effectively resolving the non-stationarity of the environment. Instead of reacting to a chaotic, unexplained price signal, an agent perceives a correlated context (e.g., high price and high congestion). This allows for the emergence of cooperation (e.g., to reduce load when congestion is high) purely from local learning, bridging the gap between the privacy of DTDE and the coordination of CTDE \cite{charbonnier2025}.

\begin{table}[h]
\centering
\caption{Technical taxonomy of MARL training paradigms in LEMs.}
\label{tab:training-taxonomy}
\resizebox{\textwidth}{!}{%
\begin{tabular}{@{}lllll@{}}
\toprule
\multicolumn{1}{c}{\textbf{Feature}} & \multicolumn{1}{c}{\textbf{CTCE}} & \multicolumn{1}{c}{\textbf{CTDE}} & \multicolumn{1}{c}{\textbf{DTDE}} & \multicolumn{1}{c}{\textbf{Ref}} \\ \midrule

Architecture & 1-to-N & N-to-N (centralized critic) & N-to-N (independent) & \cite{qiu2021a, musca2025, gronauer2022} \\
Information (Training) & Global state & Global state (centralized critic) & Local observations & \cite{wilk2024, charbonnier2025} \\
Information (Execution) & Global state & Local observations & Local observations & \cite{charbonnier2025} \\ 
Coordination mechanism & Centralized optimizer & Centralized critic & Independent learners & \cite{wilk2024, charbonnier2025} \\
Strength & Optimality & Performance & Scalability & \cite{wilk2024, charbonnier2025} \\
Weakness & Curse of dimensionality & Curse of dimensionality & Non-stationarity & \cite{qi, musca2025, gronauer2022} \\ \midrule

Optimality & High & Medium & Low & \cite{hu2024, ye2025} \\
Coordination & High & Medium & Low & \cite{charbonnier2022a} \\
Communication overhead & High & Medium & Low & \cite{hu2024, ye2025} \\
Privacy preservation & Low & Medium & High & \cite{kerena, ikeda2022, yang2025, donatus2025} \\
Scalability & Low & Low & High & \cite{charbonnier2022a, kerena, qi, yang2025} \\
Non-Stationarity Handling & High & Medium & Low & \cite{wilk2024, qi, yadav2023} \\
Trade-off & Computational complexity vs. optimality & Privact vs. performance & Non-stationarity vs. scalability & - \\ \midrule

RL algorithms & Single-agent DQN, PPO & MADDPG, MAPPO, QMIX, MATD3 & IQL, IDDPG, IPPO, Hysteretic Q & \cite{charbonnier2022a, hady2025, wilk2024, charbonnier2025} \\ 
Optimization model & MILP & ADMM & Heuristics/Local control & \cite{dvorkin2018, wuijts2023} \\ 
Game theory model & Stackelberg leader & Hybrid DRL/EGT & Independent agents & \cite{gao2025a, cheng2025} \\ \bottomrule
\end{tabular}%
}
\end{table}

\subsection{MARL algorithms for control in continuous spaces}
\label{2-sec:2.4-marl-algorithms}

As LEMs evolve from centralized, top-down dispatch models to P2P and community-based trading structures, the decision-making paradigm has shifted from static optimization to dynamic, sequential decision-making processes under uncertainty \cite{schulman2017a}. In this environment, MARL has become a computational framework that enables autonomous agents to learn optimal bidding and dispatch policies through interaction with the environment and with each other. Unlike discrete action spaces, which simplify decision-making in finite options (e.g., load, unload, or hold), continuous control enables precise modulation of active power, pricing, and energy scheduling. This granularity is essential to meet the stringent balancing requirements of day-ahead and intraday markets, where deviations in power injection or withdrawal can lead to grid instabilities or voltage violations.

\subsubsection{Comparative analysis between optimization, game theory, and RL}

To justify the selection of MARL for modern LEMs, we first analyze the limitations of traditional control paradigms, specifically optimization-based and game-theory models, which are the preferred methods for power system analysis. A comparative overview of these paradigms is presented in Table \ref{tab:lem-methods}.

Optimization-based models, such as MILP or convex optimization, have been the standard for dispatch problems. These models typically operate under the assumption of a central entity with perfect foresight and complete access to system information \cite{guerrero2019}. Their primary objective is usually the maximization of social welfare or the minimization of total system costs, implicitly assuming that all market participants act cooperatively to achieve a global optimum \cite{guerrero2020}. However, this system optimum approach fails to capture the behavioral representation of deregulated markets, where participants are self-interested agents seeking to maximize their individual benefits rather than system efficiency \cite{baringo2019, ghenai2022}.

From a modeling perspective, optimization approaches face scalability and fidelity challenges. Many studies employ the copper plate assumption, which treats the network as having infinite capacity, thus ignoring critical network constraints and leading to potentially unfeasible dispatch decisions in congested networks \cite{guerrero2019}. Furthermore, the assumption of a perfect forecast regarding the demand and generation of renewable energy is becoming increasingly unsustainable. While stochastic or robust optimization can mitigate this by incorporating sets or scenarios of uncertainty, these extensions often result in computationally complex models that produce overly conservative solutions \cite{guerrero2020, guerrero2021}. Fundamentally, they remain static solvers since they do not allow agents to adapt their behavior in real time to changing market policies or evolving competitor strategies \cite{guerrero2019}.

Game-theoretic models (e.g., Stackelberg games or Nash equilibrium analysis) address the strategic deficit of pure optimization by explicitly modeling the interactions between competitive agents. These approaches provide a theoretical basis for understanding market dynamics and identifying stable operating points where no agent has an incentive to deviate \cite{paudel2019}. However, they are often limited by strong assumptions of perfect rationality and complete information, such as that agents possess accurate models of their competitors' cost functions and constraints \cite{paudel2019}. In privacy-preserving LEMs, such information is rarely available. Moreover, finding equilibria in complex, dynamic markets with continuous action spaces presents a significant dimensionality challenge. The computational complexity of these models often renders them intractable for large-scale simulations or real-time control \cite{paudel2019}, limiting their utility to offline structural analysis rather than operational decision-making.

MARL bridges the gap between these paradigms by facilitating decentralized coordination without requiring explicit models of uncertainty or agent dynamics. Unlike robust optimization, which requires a predefined set of uncertainties, MARL agents use a representation of uncertainty, integrating the stochastic nature of renewable energy demand and generation directly into the hidden layers of the neural network through repeated interaction with the environment \cite{charbonnier2025}. This enables robust performance in the face of forecast errors without the computational explosion of scenario-based trees \cite{charbonnier2025}. Furthermore, unlike the static nature of optimization, MARL is intrinsically designed to handle non-stationarity. In a multi-agent environment, the environment changes as other agents update their policies, constituting a violation of the Markov property that breaks standard single-agent learning \cite{harder2023}. MARL algorithms, particularly those employing CTDE, address this problem by allowing agents to learn coordinated strategies using global information during training, while executing based solely on local observations \cite{harder2023a}.

\begin{table}[h]
    \centering
    \caption{Comparative analysis of market modeling and control methods in LEMs.}
    \label{tab:lem-methods}
    \resizebox{\textwidth}{!}{%
    \begin{tabular}{@{}llll@{}}
    \toprule
    \textbf{Feature} & \textbf{Optimization Models} & \textbf{Game Theoretic Models} & \textbf{MARL} \\ \midrule
    Objective & System-optimal (e.g., global welfare) & Nash equilibrium & Adaptive profit maximization \\
    Uncertainty handling & Robust or stochastic & Deterministic & Belief state (experience-based) \\
    Market dynamics & Static with perfect competition & Static equilibrium & Non-stationary and emergent \\
    Scalability & High (linear/convex) & Low (combinatorial) & High (observations and reward shaping) \\
    Limitation & Requires perfect foresight & Computational intractability & Non-stationarity and training stability \\ \bottomrule
    \end{tabular}%
    }
\end{table}

\subsubsection{PPO}

PPO has become established as a fundamental algorithm in RL applied to electrical systems. Introduced as a computationally more efficient alternative to trust region policy optimization, PPO simplifies the optimization process while preserving the advantages of trust region methods, primarily the prevention of destructive policy updates that could destabilize the learning process. In the context of LEMs, where agents must manage non-stationary environments driven by fluctuating renewable generation and dynamic pricing, the stability offered by PPO is crucial \cite{schulman2017a}.

PPO operates as an on-policy gradient method, optimizing policy based directly on data collected by the current policy. Its main innovation lies in the clipped surrogate objective function, which limits the magnitude of policy updates. This mechanism is mathematically represented by a clipped ratio that limits the deviation of the new policy from the old one in a single update step \cite{schulman2017a}. This clipping mechanism acts as a regulator in energy applications. In power systems, the reward landscape is often highly non-convex and riddled with regions where a small policy change (e.g., a slight increase in power injection) can trigger severe penalties due to constraint violations, such as line overloads or voltage spikes. Standard policy gradient methods risk policy collapse in these regions, whereas the PPO's clipped objective function avoids such drastic changes, ensuring monotonically improved policy \cite{rizki2025}.

The application of PPO in LEM has demonstrated advantages over traditional optimization methods such as MILP, specially when computational speed and adaptability are prioritized. Research indicates that PPO-based controllers can achieve near-optimal performance in economic dispatch strategies while significantly reducing the computational load associated with solving complex optimization problems at each time step. For example, in home energy management systems integrating electric vehicles and photovoltaic units, PPO has been shown to reduce daily energy costs by approximately 54\% compared to conventional scenarios, effectively learning complex, nonlinear relationships between stochastic inputs and optimal control actions without requiring an explicit model of the environment dynamics \cite{alonso2023}. Furthermore, the effectiveness of PPO extends to the presence of battery energy storage systems in joint markets involving both spot energy trading and frequency regulation. In these scenarios, the agent must continuously decide how much capacity to allocate to arbitrage versus regulation, a problem requiring continuous and precise monitoring. Research have confirmed that PPO agents can learn joint bidding strategies that significantly outperform single-market strategies, thereby maximizing total profit under uncertain market conditions \cite{anwar2022}.

A crucial comparative advantage of the PPO is its stability during the training phase. Compared to off-policy algorithms such as DDPG or TD3, the PPO exhibits lower variance in reward convergence and a robust ability to mitigate power imbalances \cite{lou2025}. Nevertheless, this conservative nature can lead to sensitivity in hyperparameter tuning. Comparative studies on power system scheduling have highlighted that the PPO may require significantly higher penalty coefficients to strictly enforce power balance constraints compared to its off-policy counterparts \cite{lou2025}. Furthermore, as an on-policy algorithm, PPO is generally less efficient in sampling, as it discards experience data after a policy update. In a LEM context where high-fidelity simulations are computationally expensive, this inefficiency can become a bottleneck \cite{rizki2025}.

\subsubsection{Asynchronous PPO}

As energy systems scale to include thousands of prosumers, the computational demands for training MARL agents increase exponentially. Asynchronous PPO (APPO) addresses the scalability limitations of standard PPO by decoupling the data collection process from the policy optimization process. This architecture is particularly relevant for LEMs modeled like Dec-POMDP, where agents operate based on local observations and require efficient mechanisms to explore large state-action spaces \cite{rigby}.

APPO introduces a high-performance architecture that facilitates distributed learning. Unlike PPO, APPO allows multiple agents to interact with their environment instances asynchronously, sending the collected trajectories to a central learner. This asynchronous sampling introduces a degree of policy lag because the behavior policy can lag behind the target policy. To correct this and ensure mathematical rigor, APPO employs importance sampling techniques, specifically V-trace, which estimates the value function and corrects the policy gradient by weighting updates based on the likelihood ratio between policies \cite{rigby}.

In the context of power systems, this architecture enables parallel simulation of diverse operating conditions. Different workers can simultaneously simulate different days, weather patterns, network topologies, or agent configurations, significantly reducing the clock time required to train complex multi-agent policies \cite{hasheminasab2025}. This capability has been successfully applied to autonomous P2P trading in networked microgrids. By modeling the trading process as a partially observable Markov game, APPO agents were able to operate effectively in the competitive environment, optimizing economic objectives while respecting transmission losses and network constraints \cite{foroughi2023}. The asynchronous nature of APPO is advantageous in this case, as it avoids the artificial lock-step synchronization typically found in simulations, allowing agents to operate on slightly different timescales that reflect the true heterogeneity of real-world prosumers \cite{foroughi2023}.

\subsubsection{SAC}

SAC represents a paradigm shift in the continuous control of energy systems by incorporating the maximum entropy RL framework. Unlike PPO, which seeks to maximize only the expected cumulative reward, SAC optimizes a dual objective: maximizing the expected reward plus the policy entropy. This entropy term incentivizes the agent to explore the action space more deeply and prevents the policy from prematurely converging to deterministic and suboptimal behavior \cite{haarnoja2018}.

The defining feature of SAC is the entropy regularization term, where a temperature parameter dynamically controls the balance between exploitation and exploration. In the context of LEMs, this mechanism is valuable for managing the multimodal nature of optimal strategies. For example, in a bidding scenario, multiple valid pricing strategies could generate similar returns; a maximum entropy policy learns a distribution across these strategies rather than narrowing down to a single point. This provides robustness against changes in the opponent's strategy, as the agent maintains a portfolio of potential optimal actions \cite{li2022}. Furthermore, as an off-policy algorithm that uses a replay buffer, SAC is significantly more sampling-efficient than PPO, making it preferable for data-scarce or slow simulation applications \cite{cheruiyot2025}.

Theoretical stability in SAC is achieved through clipped double Q-learning, which mitigates the overestimation bias common in value-based methods. This prevents the propagation of optimistic errors that could lead to aggressive and unsafe bidding behavior, a crucial feature for energy dispatch problems \cite{chao2024}. In multi-agent environments, the maximum entropy framework offers clear advantages for implicit cooperation. In a Dec-POMDP LEM, the environment is inherently non-stationary because other agents are learning simultaneously. A deterministic policy (such as those in DDPG) can be brittle; if neighbors slightly change their behavior, the optimal response could vary drastically. SAC's stochastic policy smooths the interaction landscape, allowing the agent to maintain its performance even when market dynamics drift \cite{haarnoja2018}.

In direct comparison with PPO, SAC typically demonstrates superior performance in complex tasks requiring significant exploration. For example, in building energy management tasks involving HVAC control, SAC has achieved greater energy savings and better compliance with thermal comfort standards compared to PPO, as the entropy term prevents the agent from becoming stuck in local optima \cite{sun2025}. Furthermore, in multi-microgrid scheduling, SAC has proven superior to PPO in terms of generalization and convergence capabilities for reducing operating costs \cite{gao2023}. However, the drawback is that PPO can sometimes achieve slightly lower final operating costs in static environments due to its conservative and monotonous updates \cite{khaskheli2024}. Therefore, the choice between PPO and SAC often depends on the specific constraints of the LEM. PPO is preferred for hard physical security and stability, while SAC is preferred for soft economic optimality, exploration, and robustness in highly dynamic and competitive environments. Table \ref{tab:rl-algos} presents a summary comparison of these algorithmic characteristics.

\begin{table}[h]
    {\centering
    \caption{Comparative analysis of PPO/APPO and SAC for continuous control in LEMS.}
    \label{tab:rl-algos}
    \resizebox{\textwidth}{!}{%
    \begin{tabular}{@{}llll@{}}
    \toprule
    \textbf{Feature} & \textbf{PPO / APPO} & \textbf{SAC} & \textbf{Ref} \\ \midrule
    Policy type & On-policy & Off-policy & \cite{schulman2017a, haarnoja2018} \\
    Key feature & Clip function & Max entropy & \cite{liu2023a} \\
    Trade-off & Stability / Sample efficiency & Sample efficiency / Stale data & - \\
    Sample efficiency & Low (discards data after update) & High (re-uses data via replay buffer) & \cite{schulman2017a} \\
    Update mechanism & Clipped surrogate objective & Actor-critic with entropy regularization & \cite{schulman2017a} \\
    Data sensitivity & Robust to stale data & Prone to stale data issues & \cite{sun2020} \\
    Variance/Bias & Low bias / GAE$^*$-reduced variance & Low variance / Potential off-policy bias & \cite{schulman2017a, demol2025} \\
    Strength & Stability and monotonic improvement & Exploration and sample efficiency & \cite{xu2022} \\
    LEM Suitability & Safety-critical control (e.g., voltage regulation) & Dynamic pricing and competitive bidding & \cite{demol2025} \\ \bottomrule
    \end{tabular}%
    }}
    \scriptsize{* GAE: Generalized advantage estimation.}
\end{table}

\subsection{Research gap and contribution opportunity}
\label{2-sec:2.5-XXX}

A review of the state of the art reveals a disconnect between the operational requirements of future power grids and the capabilities of existing coordination architectures. While the physical imperative to integrate DERs is evident and the theoretical promise of decentralized control is well established, the engineering challenge of achieving grid balance without relying on widespread communication or centralized authority remains an open problem.

\subsubsection{Failure of coordination for grid balance}

The primary research gap identified is the lack of a scalable, privacy-preserving mechanism for balancing supply and demand that is robust to the non-stationarity of decentralized markets. Current approaches fail to resolve the coordination challenges of scalability, privacy, and optimality, specifically in the context of the physical grid balance \cite{charbonnier2022a, guo2024}. The dominant coordination strategies for energy balancing rely on explicit communication. Mediated architectures (aggregators) create single points of failure and privacy bottlenecks \cite{shafie-khah2023}, while bilateral (P2P) architectures face a scalability obstacle due to the quadratic communication overhead required for pairwise negotiation \cite{charbonnier2022a, guo2024}. These methods solve the economic clearing problem, but they do not overcome the engineering challenge of real-time physical balancing for multiple agents, as negotiation latency exceeds the network dynamics timescales \cite{alazemi2022a}.

Existing attempts at implicit coordination for energy balance rely primarily on dynamic pricing (transactive energy) as the sole control signal \cite{ramachandran2017}. However, the literature confirms that applying price signals to automated cyber-physical systems induces system instabilities, specifically the price/load volatility and the massive load synchronization \cite{holmberg2019, tchuisseu2017}. This reveals a critical gap: no established engineering framework exists for implicit coordination that balances the electrical grid without inducing herd behavior. The field lacks a non-price-based, stigmergic coordination mechanism capable of guiding agent behavior toward physical equilibrium without the instability of purely financial signals.

The engineering challenge is more acute for medium-scale agents (commercial buildings, industrial microgrids). Unlike small residential loads, these agents possess market power as their individual actions significantly impact the state of the local grid and price formation \cite{lu2024}. Current literature focuses primarily on micro-scale (households) or macro-scale (TSO) coordination, leaving a gap in understanding how medium-scale agents can implicitly contribute to energy balance without exploiting their market power to the detriment of grid stability \cite{li2024}.

While MARL is recognized as the appropriate computational method for modeling adaptive agents, its application to the energy balance problem has presented methodological shortcomings regarding decentralization. Most applied research in MARL is based on CTDE paradigms (e.g., MADDPG) \cite{charbonnier2022a}. While these algorithms successfully learn coordination in simulation, they do so by bypassing the privacy constraint, relying on a centralized critic that accesses the global state (including the agent's private data) during training \cite{wilk2024}. This does not solve the problem of true implicit cooperation; it simply shifts centralization from execution to training. There is a clear lack of studies validating whether agents can learn to balance the network using DTDE, relying solely on local observations and engineered environmental signals (KPIs) instead of global critics \cite{hady2025}.

This research addresses these methodological and engineering shortcomings by proposing and validating a novel implicit cooperation model for energy balance in LEM. The work has the following three main contributions:

\begin{itemize}
    \item We propose a mechanism that resolves the coordination instability of price-based systems by introducing multidimensional stigmergic signals. Instead of relying solely on price, the proposed model uses reputation scores and system-level KPIs (e.g., congestion indicators, social welfare) as continuous environmental variables \cite{schubotz2022}. By integrating these metrics into the agent's observation space, we design a feedback loop where agents internalize their impact on the grid. This transforms the problem from reacting to price to maintaining system health, enabling stable energy balance as an emergent property of the system \cite{roca2016}.

    \item This research provides a systematic assessment of implicit cooperation under strict constraints of scarce information sharing. By comparing the CTCE, CTDE, and DTDE paradigms, we test the hypothesis that engineered observation spaces (stigmergy) can replace centralized critical ones. This validates whether fully decentralized agents can achieve an energy balance comparable to that of centralized baselines, demonstrating the feasibility of the proposed engineered solution in privacy-important markets \cite{wilk2024}.

\item Leveraging the MARLEM simulation framework developed in \cite{salazar2026}, this research validates coordination in a physically constrained environment. It focuses specifically on the medium-scale agent gap, studying how continuous control policies (learned using MARL for exploration) can manage the balance between maximizing individual benefits and collective grid balance \cite{zhu2022}.
\end{itemize}

\section{Methodology}
\label{2-sec:3-methodology}

The simulation framework detailed in this paper is a comprehensive, open-source tool developed for the analysis of LEMs within a MARL paradigm. A design principle of the framework is its implementation as a multi-agent environment conforming to the Gymnasium standard. This adherence ensures compatibility with an extensive range of state-of-the-art RL libraries and algorithms, thereby lowering the barrier to entry for researchers. By adopting this interface, the framework facilitates reproducible experiments and allows for the standardized benchmarking of new MARL strategies against established baselines.

To model the challenges of decision-making under uncertainty and limited information inherent in decentralized systems, the problem is formally structured as a Dec-POMDP. This formalization provides the mathematical foundation upon which the interactions between agents and their environment are built. The Dec-POMDP is defined by the tuple $\langle I, \mathcal{S}, A, 
\mathcal{P}, R, \Omega, O \rangle$, where $I$ denotes the set of agents participating in the market, $\mathcal{S}$ the set of environment states, $A$ the joint action space, $\mathcal{P}$ the state transition function, $R$ the reward function, $\Omega$ the joint observation space, and $O$ the observation function. Subindex $i$ denotes the state, space, or function for agent $i \in I$. The complete Dec-POMDP formulation is detailed in \cite{salazar2026}.

In a realistic LEM environment, no single agent has access to the global environment state $s \in \mathcal{S}$ at any given trading period $t$. Instead, agents receive local observations $o_i \in \Omega_i$ that are probabilistically correlated with the state but do not fully reveal it. This partial observability necessitates that agents maintain a belief state (i.e., a probability distribution over possible global environment states); as is common in RL, neural networks are used to encode the history of observations into a hidden state that serves as a proxy for the true state.

To facilitate learning in a partially observable environment, each agent $i \in I$ receives a local observation vector $o_i \in \Omega_i$ that constitutes a structured subset of the global environment state. It provides a sufficiently rich signal for decision-making without violating the principles of decentralization and privacy. The observation vector (see Table \ref{tab:obs-vector}) includes:

\begin{itemize}
    \item \textbf{Market signals:} Publicly available information accessible to any market participant. This encompasses the last clearing price and volume, anonymized statistics regarding P2P versus DSO trading volumes, the prevailing DSO buy (feed-in tariff) and sell (utility) prices, and the reputation score of each agent. These signals provide a common ground of information, reducing uncertainty and allowing agents to form shared expectations about the market's state.  

    \item \textbf{Agent-specific signals:} The agent's private information, which remains unobservable to other agents, thereby preserving privacy. This set of signals includes its current energy generation and demand forecasts, the state of charge (SoC) of its battery, its cumulative profit, and its dynamic reputation score. This private information is essential for the agent to tailor its strategy to its own specific circumstances and constraints.

    \item \textbf{Implicit cooperation KPIs:} A selection of system-level KPIs, such as social welfare, grid congestion, and supply-demand imbalance, that augments the observation vector. These KPIs work as a shared public signal that provides all agents with a consistent representation of the overall market health, thereby allowing them to learn the correlation between their local actions and desirable global outcomes, even without direct communication. See Section \ref{2-sec:3.2-implicit-cooperation-kpis} where a complete description of the KPIs is presented.
\end{itemize}

The objective within this environment is to find a joint policy $\pi$ that maximizes the expected discounted sum of rewards $R_i$, which in an LEM context typically correlates to net profit or utility. The complexity of solving this Dec-POMDP lies in the fact that the optimal policy for agent $i$ depends on the policies of all other agents $j \neq i$, which are evolving simultaneously. This interdependence creates a non-stationary environment for the learning algorithm, reinforcing the need for the robust, specialized MARL algorithms supported by this framework to ensure stable convergence in continuous, high-dimensional spaces.

\begin{table}[h]
\centering
\caption{Observation space vector for the formulation of the LEM as a Dec-POMDP.}
\label{tab:obs-vector}
\resizebox{\textwidth}{!}{%
\begin{tabular}{@{}cccccccc@{}}
\toprule
\multicolumn{2}{c}{\textbf{Market Signals}} & \textbf{} & \multicolumn{2}{c}{\textbf{Agent-Specific Signals}} & \textbf{} & \multicolumn{2}{c}{\textbf{Implicit Cooperation KPIs}} \\ \midrule
\begin{tabular}[c]{@{}c@{}}1. Current step\\ 2. Time of day\\ 3. Clearing price\\ 4. Clearing volume\\ 5. Grid balance\\ 6. DSO buy volume\\ 7. DSO sell volume\\ 8. DSO total volume\end{tabular} & \begin{tabular}[c]{@{}c@{}}9. P2P volume\\ 10. DSO trade ratio\\ 11. Net grid import\\ 12. DSO buy price\\ 13. DSO sell price\\ 14. Mean local price\\ 15. Price spread\\ 16. Local price advantage\end{tabular} &  & \begin{tabular}[c]{@{}c@{}}1. Energy generation\\ 2. Energy demand\\ 3. Cumulative demand satisfied\\ 4. Cumulative demand deferred\\ 5. Remaining demand\\ 6. Cummulative supply satisfied\\ 7. Cummulative supply deferred\\ 8. Remaining supply\end{tabular} & \begin{tabular}[c]{@{}c@{}}9. Mean profit\\ 10. Reputation\\ 11. Battery energy level\\ 12. Battery SoC\\ 13. Battery available charge\\ 14. Battery available discharge\\ 15. Battery cummulative charge\\ 16. Battery cummulative discharge\end{tabular} &  & \begin{tabular}[c]{@{}c@{}}1. Social welfare\\ 2. Market liquidity\\ 3. Mean bid-ask spread\\ 4. Price volatility\\ 5. Supply-demand imbalance\end{tabular} & \begin{tabular}[c]{@{}c@{}}6. Grid congestion\\ 7. Coordination score\\ 8. Coordination convergence\\ 9. DER self-consumption\\ 10. Flexibility utilization\end{tabular} \\ \bottomrule
\end{tabular}%
}
\end{table}

\subsection{Implicit cooperation model}
\label{2-sec:3.1-framework-architecture}

The implicit cooperation model represents a fundamental departure from traditional centralized control and explicit negotiation paradigms. Building upon the MARLEM simulation framework established in our previous work \cite{salazar2026}, which provided the physical and market layers of the environment, this section details the specific algorithmic contributions designed to enable coordination. Unlike explicit coordination mechanisms that require direct communication, centralized orchestration, or predefined protocols where agents must share information about their intentions and capabilities \cite{schubotz2022}, implicit cooperation emerges naturally through the strategic design of information structures and incentive mechanisms. In our framework, agents learn to coordinate their actions without explicit communication, relying instead on system-level signals that reflect the collective state of the market and grid.

We define \emph{implicit cooperation} in the context of decentralized energy markets as the emergence of coordinated behaviors among self-interested agents that simultaneously satisfy individual economic objectives and collective grid constraints. This is achieved through the interaction of three components: shared information signals in the form of system-level KPIs that provide agents with information about market and grid health, incentive alignment via reward structures that link individual agent performance to system-level outcomes, and the development of complementary strategies through repeated interactions and learning.

The motivation for this approach is grounded in the theory of emergent behavior in complex systems. In decentralized systems, explicit coordination mechanisms face challenges including communication overhead and latency, privacy concerns regarding sensitive data, the risk of single points of failure, and inherent scalability limitations \cite{charbonnier2022a, guo2024}. Implicit cooperation addresses these challenges by enabling coordination through shared environmental signals rather than direct agent-to-agent communication, following principles similar to quantitative, marker-based stigmergy in swarm intelligence \cite{schubotz2022}.

In this analogy, the market platform functions as the stigmergic medium \cite{denicola2020}, and an agent's action leaves a trace (e.g., congestion signal) in the environment, which subsequently stimulates the performance of a next action by other agents. The role of system-level KPIs is critical in this framework. These indicators serve as the pheromones that allow agents to infer the global state from local observations \cite{schubotz2022}. By including KPIs regarding market efficiency, grid balance, and resource utilization in both the observation space and the reward function, agents can learn to associate their local actions with grid-level outcomes. This creates a feedback loop where self-interested learning leads to the emergence of strategies that stabilize the grid, effectively decoupling the complexity of the system from the complexity of the individual agents \cite{roca2016}.

The implementation of implicit cooperation relies on two coupled mechanisms: an augmented observation space design and a multi-objective reward function. While the complete mathematical derivation of the environment is detailed in \cite{salazar2026}, we focus here on the specific augmentations that drive cooperative behavior.

The design of the observation space is fundamental to enabling implicit cooperation under the constraints of partial observability. In a standard Dec-POMDP, an agent $i$ typically observes only its local state. To enable coordination without violating privacy, we introduce an augmented observation space $\widehat{\Omega}$. Each agent receives a local observation $o_i \in \widehat{\Omega}_i$ that includes not only its internal state (e.g., generation/demand profiles, battery SoC, profit) but also a comprehensive set of system-level signals. To guide economic strategy, the observation space includes dynamic market signals such as the clearing price, clearing volume, and the price spread. Crucially, agents also observe the P2P trade ratio, defined as the proportion of P2P trades relative to the total trade volume. A high P2P trade ratio serves as a signal of market self-sufficiency, implicitly encouraging agents to prioritize the local layer over the DSO.

Simultaneously, grid state signals are injected to internalize physical constraints. The grid balance ($B_{grid}$) acts as a normalized indicator of the net energy position of the grid (see (\ref{eq:grid-balance-2}), where $G_t$ is the energy generation, $D_t$ the energy demand, and $E_{bought, t}$ and $E_{sold, t}$ are the energy bought and sold for agent $i$ at trading period $t$), where positive values indicate a surplus and negative values a deficit. A positive $B_{grid}$ indicates a generation surplus, while a negative value indicates a deficit. This is complemented by the grid congestion metric (see (\ref{eq:grid-congestion-2}), where $\mathcal{F}$ is the power flow for each power line $e \in \xi$, $C$ is the capacity at the grid edge, and $\upzeta$ the mean congestion level), which represents the average congestion level across all network edges normalized to $[0, 1]$. Low congestion levels are indicative of a system operating well within its physical limits, enhancing reliability. These signals function as congestion indicators, warning agents of infrastructure stress directly without accessing the topology data of the grid operator \cite{deng2019}.

\begin{equation}
    \label{eq:grid-balance-2}
    B_{grid} = \sum_{i \in I}(G_i - D_i + E_{bought,i} - E_{sold,i})
\end{equation}

\begin{equation}
    \label{eq:grid-congestion-2}
    \bar{\upzeta} = \frac{1}{|\xi|} \sum_{e \in \xi} \frac{\mathcal{F}_e}{\max(C_e)}
\end{equation}

Finally, specific implicit cooperation KPIs are injected to guide high-level strategy. These KPIs are categorized to provide a multi-faceted view of system performance, and include KPIs for economic efficiency, grid stability, resource coordination, and coordination effectiveness. The complete suite of implicit cooperation KPIs is described in more detail in \cite{salazar2026}. By broadcasting these signals, the system converts global state information into local observables without revealing individual private data, thereby preserving privacy while enabling system-aware learning.

The reward function is the primary driver of behavioral adaptation and is designed to balance individual economic incentives with system-level stability goals. As derived in \cite{salazar2026}, the reward $R_{i}$ for agent $i$ at trading period $t$ is a composite function creating a hierarchical incentive mechanism (see (\ref{eq:reward-function-2})).

\begin{equation}
\label{eq:reward-function-2}
    R_i = R_{base,i} \cdot (1 + f_{coop} \cdot f_{contrib,i}) - \gamma_{DSO,i} - \gamma_{UD,i}
\end{equation}

The base reward ($R_{base,i}$) is a weighted sum that represents the individual performance of the agent and ensures that agents learn effective bidding strategies to trade on their own behalf. $R_{base,i}$ comprises an economic component which rewards profit maximization relative to DSO prices and incentivizes agents to learn effective bidding strategies, a grid balance component which rewards agents for actions that help reduce overall grid imbalance (i.e., buying during surplus periods and selling during deficit periods), a resource allocation component which encourages efficient utilization of available DER capacity, a trading component which incentivizes successful transactions with a higher weight for P2P transactions to create a preference for local coordination, and a stability component which rewards behaviors that contribute to long-term market stability.

The core contribution of this work regarding implicit cooperation is the cooperation bonus, represented by the multiplicative term involving the cooperation factor $f_{coop}$, which is calculated from the weighted sum of normalized system KPIs (including economic efficiency, grid stability, and coordination effectiveness). It acts as a multiplicative bonus that scales the base reward. When the system is healthy (i.e., high social welfare and low congestion) all agents perceive higher potential rewards. This creates a dynamic where agents are motivated to maintain a stable market environment in order to maximize their own long-term utility \cite{denicola2020, schraudner2021}. To prevent reward exploitation, the cooperation bonus is modulated by the contribution factor $f_{contrib,i}$, which quantifies agent $i$'s specific individual impact on the system health. $f_{contrib,i}$ is a weighted sum of the agent's marginal contribution to reducing grid imbalance ($-\frac{\partial |B_{grid}|}{\partial a_i}$), its contribution to price efficiency, and its contribution to local trading volume. This term ensures effective credit assignment, a challenge in cooperative MARL, by rewarding agents proportional to their positive impact on the collective state. Finally, the structure includes a penalty mechanism, encouraging agents to prioritize P2P transactions and fostering a self-sufficient local market. $\gamma_{DSO,i}$ is applied for trading with the DSO, scaled by the current grid imbalance, and $\gamma_{UD,i}$ is applied for the unmet demand. This creates a dual incentive: agents are discouraged from relying on the DSO, and this discouragement is amplified during critical imbalance periods, forcing agents to seek local matching solutions.

These mechanisms establish a continuous feedback loop of measurement, sharing, and adaptation. First, the system calculates all KPIs based on the current market state and grid conditions. These KPIs are then integrated into each agent's augmented observation space, providing a shared signal about system health. Simultaneously, these KPIs determine the magnitude of the cooperation factor in the reward function. As agents execute actions to maximize this reward, they learn to associate specific behaviors (such as reducing load when the coordination score is low) with system-level outcomes. This feedback loop enables agents to learn coordination strategies without explicit communication, as they can infer the impact of their actions on the collective through the KPI signals.

\subsection{Implicit cooperation KPIs}
\label{2-sec:3.2-implicit-cooperation-kpis}

The framework calculates a comprehensive suite of KPIs, categorized to provide a multi-faceted view of system performance \cite{salazar2026}. The following are the indicators used to measure:

\begin{enumerate}
    \item \textbf{Economic efficiency KPIs:} This set of metrics measures the market's ability to create value and facilitate efficient price discovery.
    
    \begin{itemize}
        \item \textbf{Social welfare:} The total economic value of all trades, representing the sum of consumer and producer surplus. It serves as the primary indicator of overall market efficiency. Note that $p$ is the price and $q$ the energy quantity.

        \begin{equation}
            \label{eq:social-welfare-2}
            \text{Social Welfare} = \sum_{trades} p_{trade} \cdot q_{trade}
        \end{equation}

        \item \textbf{Market liquidity:} The total volume of energy traded, indicating market activity and depth.

        \begin{equation}
            \label{eq:market-liquidity-2}
            \text{Liquidity} = \sum_{trades} q_{trade}
        \end{equation}

        \item \textbf{Average bid-ask spread:} The average difference between buy and sell order prices, measuring market efficiency.

        \begin{equation}
            \label{eq:bid-ask-spread-2}
            \text{Spread} = \mathbb{E}[p_{ask}] - \mathbb{E}[p_{bid}]
        \end{equation}

        \item \textbf{Price volatility:} The standard deviation of the clearing price over a time window, indicating market stability.
    \end{itemize}

    \item \textbf{Grid stability KPIs:} This set of metrics assesses the physical health and operational efficiency of the electrical grid.
    
    \begin{itemize}
        \item \textbf{Supply-demand imbalance:} The net energy imbalance normalized by grid capacity $C_{grid}$, measuring how well supply and demand are balanced. A value near zero suggests stable operation that does not strain the wider grid.

        \begin{equation}
            \label{eq:imbalance-2}
            \text{Imbalance} = \frac{| \sum q_{buy} - \sum q_{sell} |}{C_{grid}}
        \end{equation}

        \item \textbf{Grid gongestion:} The average congestion level across all power lines $e \in \xi$ as defined in (\ref{eq:grid-congestion-2}), indicating physical stress on the infrastructure. Low congestion levels are indicative of a system operating well within its physical limits, enhancing reliability.

        \item \textbf{Grid balance:} The overall energy balance of the grid, calculated as the difference between total generation and consumption as in (\ref{eq:grid-balance-2}).
    \end{itemize}

    \item \textbf{Resource coordination KPIs:} These metrics evaluate how effectively DERs are utilized and coordinated.
    
    \begin{itemize}
        \item \textbf{DER self-consumption:} The proportion of total energy transacted that occurs in P2P trades as opposed to with the DSO. A high value is indicative of a self-sufficient and effective local market, reducing reliance on centralized utilities.

        \begin{equation}
            \label{eq:self-consumption-2}
            \text{Self-Consumption} = \frac{\sum q_{P2P}}{\sum (q_{P2P} + q_{DSO})}
        \end{equation}

        \item \textbf{Flexibility utilization:} The proportion of available flexible energy that is actively utilized in P2P trading. This metric measures how effectively agents utilize their energy flexibility resources (generation surplus, demand deficit, and battery capacity).

        \begin{equation}
            \label{eq:flexibility-2}
            \text{Flexibility Utilization} = \frac{\sum q_{p2p}}{\sum q_{available}}
        \end{equation}

        where $\sum q_{available}$ is the total available flexibility across all agents, considering the sellable flexibility (surplus generation plus battery discharge capacity), and the buyable flexibility (deficit demand plus battery charge capacity) at a given trading period $t \in \{1, \dots, \tau\}$.
    \end{itemize}

    \item \textbf{Coordination effectiveness KPIs:} These metrics measure the emergence and effectiveness of coordination among agents.
    
    \begin{itemize}
        \item \textbf{Coordination score:} A measure of coordination, reflecting the market's balance. A score approaching 1 indicates perfect system balance.

        \begin{equation}
            \label{eq:coordination-2}
            \text{Coordination Score} = 1 - \text{Imbalance}
        \end{equation}

        \item \textbf{Coordination convergence:} Measures the stability of trading volumes over a recent window, indicating if a stable, coordinated pattern has emerged. It is calculated similarly to the price volatility metric.
    \end{itemize}
\end{enumerate}

\subsection{MARL algorithms: algorithm-specific characteristics}
\label{2-sec:3.2-marl-algorithms}

\subsubsection{PPO}

As an on-policy gradient method, PPO learns directly from the data generated by the current policy. It was designed to address the step size problem in policy gradient methods: if a policy update is too large, the agent may move into a region of the parameter space that yields bad performance. Because on-policy agents collect data based on their own behavior, a bad policy leads to bad data, creating a terminal performance collapse from which the agent cannot recover.

The core innovation of PPO is the clipped surrogate objective function, which enforces a trust region to ensure that policy updates are incremental. For a decentralized prosumer agent $i$, the objective $J^{CLIP}(\theta_i)$ is defined in (\ref{eq:objective-func-ppo}).

\begin{equation}
\label{eq:objective-func-ppo}
    J^{CLIP}(\theta_i) = \mathbb{E}_t \left[ \min \left( r_t(\theta_i) \widehat{A}_t, \text{clip}(r_t(\theta_i), 1-\epsilon, 1+\epsilon) \widehat{A}_t \right) \right]
\end{equation}

\begin{equation}
    r_t(\theta_i) = \frac{\pi_{\theta_i}(a_t|o_t)}{\pi_{\theta_{old}}(a_t|o_t)}
\end{equation}

Where $r_t(\theta_i)$ is the probability ratio that measures how much more (or less) likely an action $a_t$ is under the new policy compared to the old , and $\theta_i$ is the policy parameters of agent $i \in I$. The term $\widehat{A}_t$ is the advantage, which quantifies how much better a specific action was compared to the average behavior. The operational logic of the clipping mechanism $(\epsilon$, typically 0.2) varies based on the nature of the experience:

\begin{itemize}
    \item \textbf{Positive advantage ($\widehat{A}_t > 0$)}: If the action was good, the algorithm wants to increase $r_t$. The $\min$ operator caps this increase at $1+\epsilon$. This prevents the policy from over-optimizing based on a single lucky market interaction, ensuring stability.

    \item \textbf{Negative advantage ($\widehat{A}_t < 0$):} If the action was bad, the algorithm wants to decrease $r_t$. The clipping prevents the probability from being driven to zero instantaneously, which would destroy the agent's ability to explore better strategies in the future.
\end{itemize}

To compute the advantage $\widehat{A}_t$, PPO utilizes the generalized advantage estimation. This mechanism reduces the variance of the rewards received from the LEM by interpolating between immediate feedback and long-term returns.

In the context of implicit cooperation, PPO's strength lies in this dampening effect. In a multi-agent environment, the transition dynamics are non-stationary because other agents are learning simultaneously. If Agent A changes its bidding strategy drastically, Agent B's environment changes. PPO’s conservative updates allow neighbors time to adapt to these shifts, facilitating convergence toward stable Nash equilibria rather than the chaotic oscillations often observed in non-clipped methods.

\subsubsection{APPO}

APPO addresses the computational bottleneck of standard PPO in medium-scale simulations. In an LEM with a hundred agents, the time required to solve market-clearing physics can vary. Standard PPO is synchronous, meaning the learner must wait for every agent to finish its simulation before updating. APPO decouples actors (data collectors) from the learner (policy optimizer) in an actor-learner topology.

Because collection and optimization happen in parallel, a policy lag occurs: the data might have been collected by a slightly older version of the policy ($\pi_{behavior}$) than the one currently being updated ($\pi_{target}$). Using this stale data introduces bias. APPO employs V-trace to correct this discrepancy mathematically. The V-trace target for value updates is defined in (\ref{eq:v-trace}), where $V_t^{target}$ is the corrected value estimate that the learner aims to reach for the state at time $t$, $V(s_t)$ is the current value function's estimate of the state $s_t \in \mathcal{S}$, $\psi$ is the discount factor which determinies how much the agent values future rewards versus immediate ones, $\ell$ is the length of the trajectory segment, $c_{t+j}$ is a trace-cutting coefficient that controls how much the errors from future steps affect the current update, $\rho_{t+k}$ is the truncated importance sampling weight (being $\bar{\rho}$ a hyperparameter representing the clipping threshold to prevent the importance weight from becoming too large), and $\delta_{t+k}$ is the temporal difference error.

\begin{equation}
\label{eq:v-trace}
    V_t^{target} = V(s_t) + \sum_{k=0}^{\ell-1} \psi^k \left(\prod_{j=0}^{k-1} c_{t+j}\right) \rho_{t+k} \delta_{t+k}
\end{equation}

\begin{equation}
    \rho_t = \min \left( \bar{\rho}, \frac{\pi_{target}(a_t|o_t)}{\pi_{behavior}(a_t|o_t)} \right)
\end{equation}

The importance weight $\rho_t$ measures how much the current policy has drifted from the one that collected the data. If the drift is too high, the weights effectively cut the trace, forcing the learner to rely more on its current value estimates rather than the stale rewards.

In our multi-agent context, APPO provides experience diversity. Because different actors can interact with different instances of the energy grid simultaneously, the learner receives a broader range of scenarios (e.g., varying generation levels across different simulated days). This prevents the lock-step synchronization of strategies and encourages the development of decentralized policies that are robust to a wide variety of grid conditions.

\subsubsection{SAC}

While PPO maximizes the expected return, SAC maximizes the expected return plus the entropy ($\mathcal{H}$) of the policy. The objective function is defined in (\ref{eq:objective-func-sac}), where $\alpha$ is the temperature parameter.

\begin{equation}
\label{eq:objective-func-sac}
    J(\pi) = \sum_{t=0}^\tau \mathbb{E}_{(o_t, a_t) \sim \rho_\pi} \left[ r(o_t, a_t) + \alpha \mathcal{H}(\pi(\cdot|o_t)) \right]
\end{equation}

Entropy is a measure of randomness. By maximizing it, SAC forces the agent to remain as stochastic as possible while still achieving the goal. This provides three critical benefits for LEM coordination:

\begin{itemize}
    \item \textbf{Robustness to non-stationarity:} In a Dec-POMDP, deterministic policies are brittle. If a neighbor changes a bidding threshold, a deterministic policy might suddenly fail. A stochastic policy learns a distribution of optimal actions, making it more resilient to the shifting behaviors of peers.

    \item \textbf{Discovery of cooperative strategies:} Coordinated load shifting is difficult to discover because it often requires a temporary local economic loss for a global system benefit. SAC’s mandated entropy prevents the agent from prematurely converging to a selfish, suboptimal local minimum strategy.

    \item \textbf{Reparameterization for continuous control:} To optimize the stochastic policy, SAC makes the sampled action $a_t$ a differentiable function of the policy parameters. This allows for the precise, continuous modulation of bid quantity and bid price bids needed for fine-grained grid balancing.
\end{itemize}

SAC is an off-policy algorithm, utilizing a replay buffer to store every past interaction. This makes it more sample-efficient than PPO. To prevent agents become overly optimistic about certain actions due to errors in the neural network, SAC maintains two independent Q-networks and uses the minimum of the two for its updates, ensuring that the coordination signals learned by the agents are grounded in stable value estimates.

\subsection{MARL training paradigms}
\label{2-sec:3.2-training-paradigms}

\subsubsection{CTCE}

CTCE represents the theoretical upper bound for coordination quality and serves as the primary benchmark for maximum achievable performance in this study. In this paradigm, the multi-agent problem is reduced to a single-agent problem with a high-dimensional joint action space. The architecture utilizes a single policy network $\pi_{\theta}$ that receives the global environment state $\mathcal{S}_t$. This global state provides a complete description of the environment, concatenating all private information from the grid. The policy maps this global state directly to the joint action vector $\mathbf{a}_t$, solving the optimization problem in a single forward pass. The mathematical formulation assumes a joint policy that, while it may be factorized for implementation, remains globally conditioned (see (\ref{eq:ctce-policy})).

\begin{equation}
\label{eq:ctce-policy}
\pi_{\theta}(a_{1,t}, \dots, a_{I,t} | s_t) = \prod_{i=1}^{I} \pi_{\theta,i}(a_{i,t} | s_t)
\end{equation}

The shared conditioning on $\mathcal{S}_t$ enables the network to learn sophisticated correlations between agent behaviors. The centralized value function $V_{\phi}(s)$ estimates the expected return by considering the aggregate system-level reward, thereby directly optimizing for the global objective $J_{CTCE}(\theta)$ defined as the expected discounted global return (see (\ref{eq:ctce-obj})).

\begin{equation}
\label{eq:ctce-obj}
J_{CTCE}(\theta) = \mathbb{E}_{s \sim \rho^\pi, \mathbf{a} \sim \pi_\theta} \left[ \sum_{t=0}^T \sum_{i=1}^I \psi^t R_{i}(s_t, \mathbf{a}_t) \right]
\end{equation}

Because the controller has complete visibility and authority, the training process is stationary; the environment dynamics are fixed, and there are no other independent learning entities to create non-stationarity. CTCE can achieve Pareto optimality by explicitly trading off the utility of one agent against another to satisfy system constraints. However, CTCE serves strictly as a benchmark due to its violation of privacy and scalability issues, as the search space volume grows exponentially with the number of agents.

\subsubsection{CTDE}

The operating logic of CTDE relies on a hybrid actor-critic architecture. During the training phase, a centralized critic $Q_{\phi}(s, \mathbf{a})$ is utilized. This critic has access to the global state $\mathcal{S}_t$ and the joint action vector $\mathbf{a}_t$, allowing it to estimate the Q-value with a full perspective of the system's coupling. However, during the execution phase, the critic is discarded, and each agent $i$ retains an independent policy network $\pi_{\theta_i}(a_i | o_i)$ (decentralized actor) that maps only its local observation $o_i$ to an action $a_i$. The gradient for updating the actor policy $\theta_i$ is computed using the deterministic policy gradient theorem or its stochastic equivalent.

\begin{equation}
\label{eq:ctde-gradient}
\nabla_{\theta_i} J(\theta_i) = \mathbb{E}_{s, \mathbf{a}} \left[ \nabla{\theta_i} \log \pi_{\theta_i}(a_i | o_i) Q_{\phi}(s, a_{1,t}, \dots, a_{I,t}) \right]
\end{equation}

In (\ref{eq:ctde-gradient}), the centralized critic evaluates the quality of agent $i$'s action $a_i$ not based on local outcomes, which might be noisy due to the actions of peers, but in the context of the global system state. This resolves the non-stationarity problem during training: the critic explicitly accounts for the shifting policies of all other agents $\pi_{-i}$, ensuring that the advantage signal remains stable. This informed signal allows the local actor to learn coordination strategies that can later be executed autonomously.

In our study, CTDE tests whether coordination strategies can be explicitly taught via centralized guidance. However, it still presents a privacy hurdle during the training phase, as agents must share their experiences (observations, actions, and rewards) with the central critic, which may conflict with data protection regulations in real-world LEMs.

\subsubsection{DTDE}

In DTDE, the multi-agent system is treated as a set of independent learners. Each agent $i$ maintains its own actor $\pi_{\theta_i}(a_i | o_i)$ and its own independent critic $V_{\phi_i}(o_i)$ or $Q_{\phi_i}(o_i, a_i)$. Critically, the critic only has access to the local observation $o_i$ and the agent's own action $a_i$, treating the actions of all other agents purely as part of the environment dynamics. The training objective for each agent is to maximize its own expected discounted return, as defined in (\ref{eq:dtde-obj}).

\begin{equation}
\label{eq:dtde-obj}
J_{DTDE}(\theta_i) = \mathbb{E}_{o_i, a_i} \left[ \sum_{t=0}^\tau \psi^t R_i(o_{i,t}, a_{i,t}) \right]
\end{equation}

The central theoretical hurdle of DTDE is non-stationarity. In a Dec-POMDP, from the perspective of agent $i$, the environment transition function $\mathcal{P}_i$ effectively absorbs the evolving policies of all other agents $\pi_{-i}$ (see (\ref{eq:non-stationarity})).

\begin{equation}
\label{eq:non-stationarity}
\mathcal{P}_i (s' | s, a_i) = \sum_{\mathbf{a}_{-i} \in \mathcal{A}_{-i}} \mathcal{P}(s' | s, a_i, \mathbf{a}_{-i}) \prod_{j \neq i} \pi_j(a_j | o_j)
\end{equation}

Because all agents update their policies $\pi_j$ simultaneously during training, $\mathcal{P}_i$ becomes time-variant. This violation of the Markov property leads to learning instability, oscillations, and convergence to suboptimal Nash equilibria.

Our contribution is to mitigate this non-stationarity through the augmented observation space described in Section \ref{2-sec:3.1-framework-architecture}. By injecting system-level KPIs into the local observation $o_i$, the environment becomes quasi-stationary. These KPIs provide a compressed, privacy-preserving representation of the aggregate behavior of the other agents. Instead of needing to know Agent B's specific state, Agent A only needs to know the physical impact of Agent B's actions on the grid.

\section{Experimental setup}
\label{2-sec:4-experimental-setup}

This case study is designed to test the implicit cooperation hypothesis under realistic conditions. Specifically, it examines whether decentralized agents can achieve system-level coordinationm such as maintaining supply-demand balance, using only local observations augmented with stigmergic KPI signals, thereby bypassing the need for centralized dispatch or explicit communication. The scenario is evaluated across all combinations of the training paradigms (CTCE, CTDE, DTDE) and MARL algorithms (PPO, APPO, SAC).

The primary goal of this design is to quantitatively verify whether decentralized agents, operating under the constraints of scarce information (DTDE), can approximate the coordination quality of theoretically optimal centralized systems (CTCE). We assess which continuous control algorithm (PPO, APPO, or SAC) best navigates the trade-off between exploration—essential for discovering cooperative equilibria—and stability, which is critical for maintaining physical grid constraints in a non-stationary environment. By subjecting diverse algorithms to identical environmental conditions (specifically an heterogeneous agent population and constrained IEEE 34-bus topology) we eliminate confounding variables. This ensures that observed performance differences are attributable solely to the interaction between the training paradigm and the learning update rule. We test the hypothesis that system-level KPIs (e.g., the Grid Balance Index) injected into the observation space function as effective proxies for centralized coordination signals, allowing agents to internalize the externalities of their actions.

Detailed configurations and implementations for replicating this case study can be found in the project's dedicated GitHub repository (\href{https://github.com/salazarna/marlem}{https://github.com/salazarna/marlem}).

\subsection{Case study configuration}
\label{2-sec:4.1-case-study-configuration}

\begin{figure}[h]
    \centering
    \includegraphics[width=\linewidth, trim=0cm 0cm 0cm 1cm, clip]{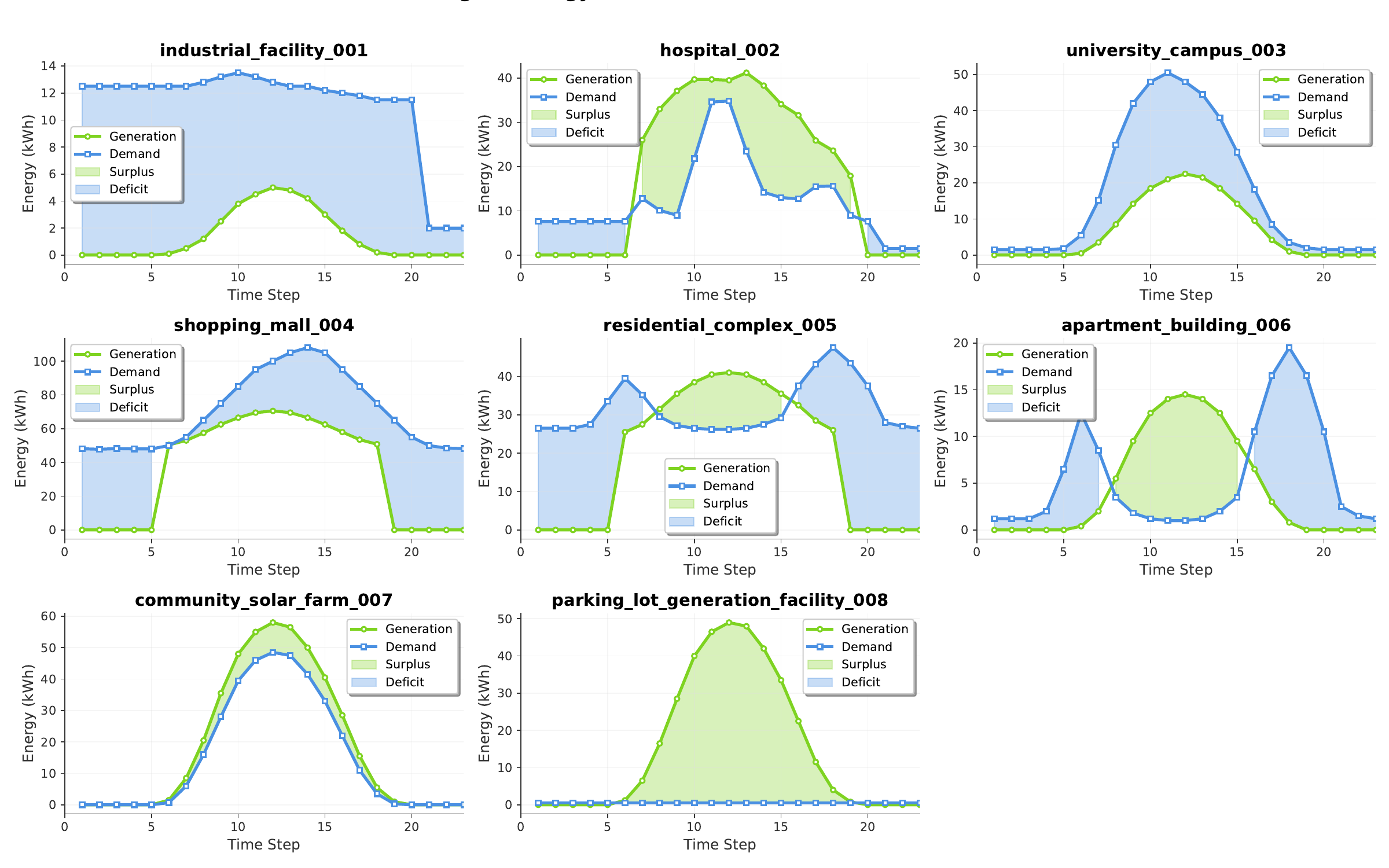}
    \caption{Generation and demand profiles for DER agents in the case study.}
    \label{fig:agents-profile}
\end{figure}

The experimental configuration enforces interdependence through a diverse population of 8 DER agents with complementary generation and demand profiles (see Fig. \ref{fig:agents-profile}). This heterogeneity ensures that coordination is not just beneficial but necessary for optimal system performance. The capacity ranges (75–180 kW) and battery ratios are selected to ensure realistic penetration levels (40-60\% of total load) that provide meaningful improvement potential without causing immediate grid instability. Table \ref{tab:agent-configuration} details the specific configuration for each agent, including their roles, capacities, and the behavioral logic underpinning their profiles. To ensure reproducibility, the environment configuration is detailed in Table \ref{tab:simulation-parameters}.

\begin{table}[h]
\centering
\caption{Configuration of the agents for the case study.}
\label{tab:agent-configuration}
\resizebox{\textwidth}{!}{%
\begin{tabular}{@{}cccccl@{}}
\toprule
\textbf{Agent} & \textbf{IEEE-34 Node} & \textbf{Capacity (kW)} & \textbf{Battery (kW)} & \textbf{Battery Ratio} & \textbf{Description} \\ \midrule
Small Industry & 840 & 180.0 & 108.0 & 0.6 & Continuous demand with morning peak; relies on stored energy \\

Community Hospital & 890 & 170.0 & 85.0 & 0.5 & High morning demand peak, transitioning to afternoon generation surplus \\

University Campus & 844 & 140.0 & 98.0 & 0.7 & Bi-modal demand (morning/afternoon peaks); moderate midday generation \\

Shopping Mall & 816 & 130.0 & 78.0 & 0.6 & High afternoon/evening demand; generation sink for midday surplus \\

Residential Complex & 800 & 80.0 & 64.0 & 0.8 & Dual-peak demand (morning/evening); reliant on storage for evening ramp \\

Apartment Building & 830 & 75.0 & 52.5 & 0.7 & Similar to residential complex; moderate generation capability \\

Community Solar Farm & 848 & 180.0 & 180.0 & 1.0 & High midday generation, minimal demand; acts as flexible provider \\

Parking Lot & 860 & 160.0 & 144.0 & 0.9 & High midday generation; acts as flexibility provider \\ \bottomrule
\end{tabular}%
}
\end{table}

Each agent is equipped with a battery modeled with a 95\% charge/discharge efficiency. To preserve battery health and realism, the SoC is constrained between 5\% and 95\% of nominal capacity. These differing battery ratios (0.5 to 1.0) create a heterogeneous landscape of market power, allowing to analyze how agents with different storage capabilities contribute to grid balance.

\begin{figure}[h]
    \centering
    \includegraphics[width=\linewidth, trim=0cm 0cm 0cm 1cm, clip]{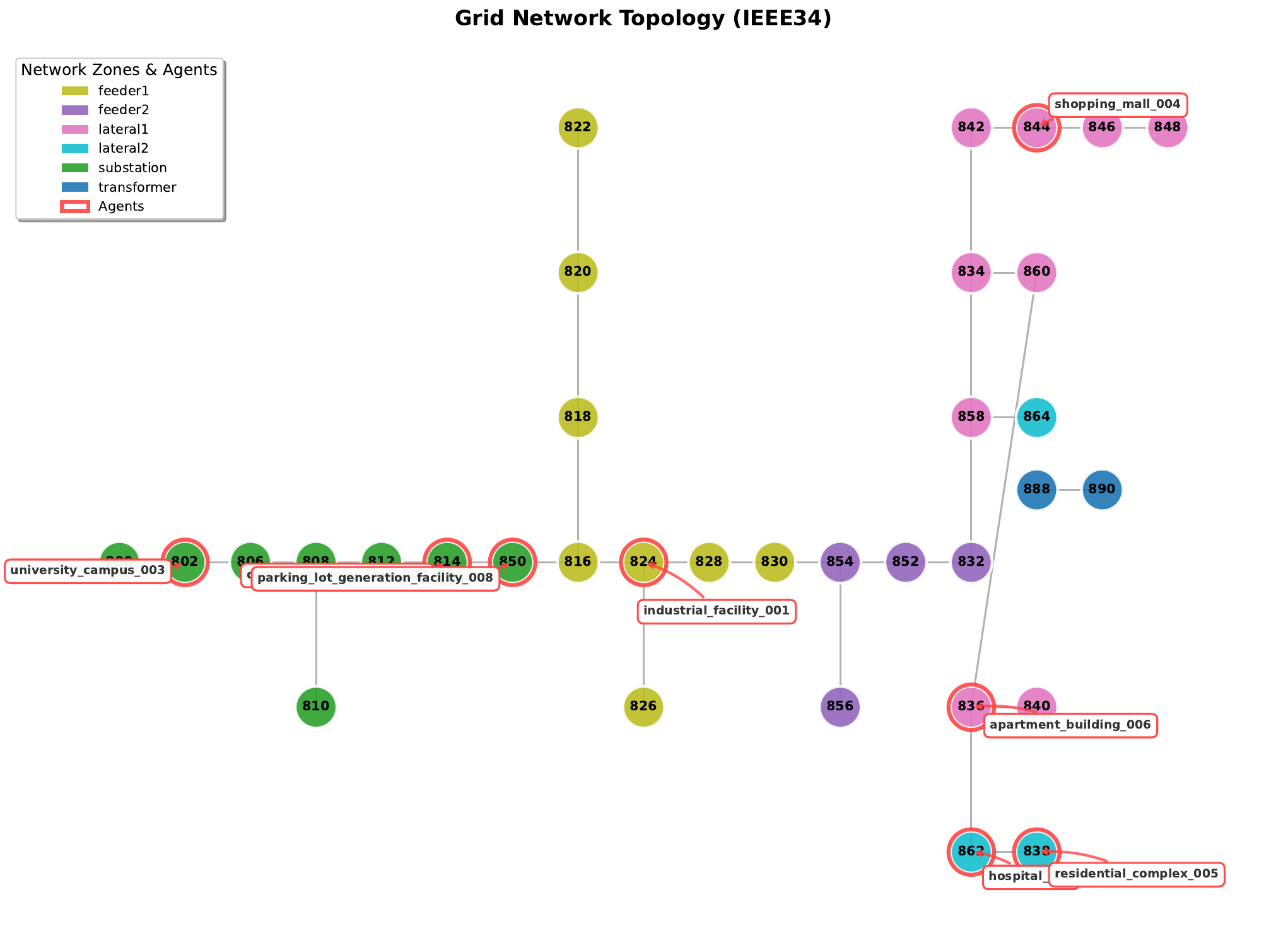}
    \caption{The 34-node IEEE test feeder as the grid network for the case study.}
    \label{fig:grid-network}
\end{figure}

The simulation operates on an IEEE 34-bus test feeder topology, a standard distribution network benchmark modified for this case study. The total system capacity is constrained to 1800 kW. This capacity limit is binding during peak usage hours (given the total agent capacity of roughly 1115 kW plus baseload), forcing agents to coordinate physically via local balancing rather than relying solely on the DSO. The strategic placement of agents (see Fig. \ref{fig:grid-network}) is determined by the topology's critical electrical points to enable robust hardware-in-the-loop validation. Agents are assigned to specific nodes that correspond to designated signal monitoring points within the future RT-Lab IEEE-34 model implementation. In the substation zone, the \emph{small industry} is located at node 800 (substation bus), which serves as the primary feed point and a critical location for measuring grid entry/exit flows. The \emph{community hospital} agent is positioned at node 890 (transformer secondary), a key measurement node for assessing power quality and voltage regulation immediately downstream of the voltage transformation. Deeper within the network, the \emph{shopping mall} is situated at node 816, a major feeder branch point selected to validate coordination dynamics at critical network bifurcations. Finally, to test system stability across significant electrical distances, the \emph{community solar farm} and \emph{parking lot} are placed at network extremities (node 860 and node 848 at the end of the first lateral, respectively) where voltage support is most critical and will be rigorously monitored during hardware-in-the-loop validation.

\begin{figure}[h]
    \centering
    \includegraphics[width=0.6\linewidth, trim=0cm 0cm 0cm 1cm, clip]{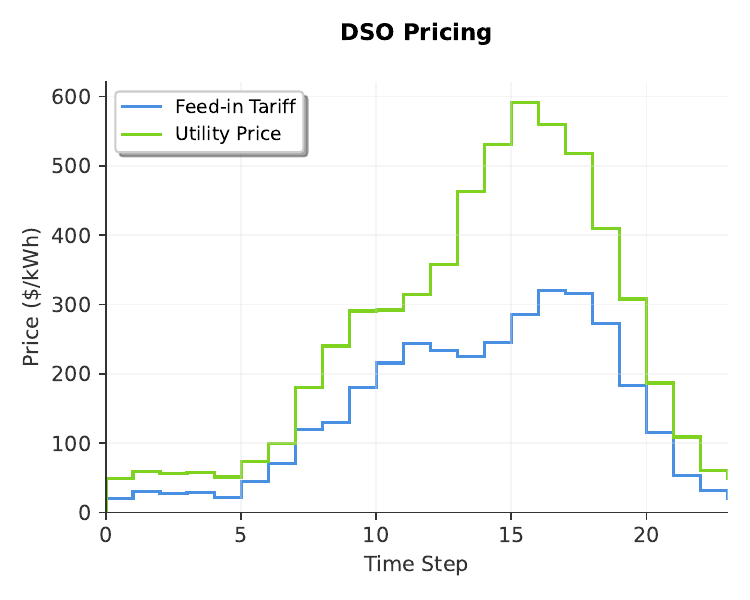}
    \caption{Feed-in tariff and utility price profiles for DSO agent in the case study.}
    \label{fig:dso-pricing}
\end{figure}

The market layer is configured to incentivize local coordination over reliance on the main grid. We employ an average pricing mechanism ($p_{trade} = \frac{p_{buy} + p_{sell}}{2}$). This mechanism is selected to promote fair price discovery and reduce the complexity of strategic bidding. The market parameters include a price floor of 20 \$/kWh and a ceiling of 600 \$/kWh to encourage local trading over DSO arbitrage. Partner preferences are enabled, allowing agents to learn and prioritize reliable peers. Crucially, the DSO price signals act as the boundary conditions for agent behavior. As illustrated in Fig. \ref{fig:dso-pricing}, a distinct gap is maintained between the feed-in tariff (FIT) and the utility price. The FIT represents the revenue for selling to the grid and it remains consistently lower than the utility price (peaking around ~300 \$/kWh). The utility price represents the cost to buy from the grid and it peaks significantly during the day (reaching ~600 \$/MWh), creating an economic penalty for agents that fail to source energy locally during peak demand. The gap between these two curves creates the P2P margin, where agents are economically incentivized to trade with each other because a seller can earn more than the FIT, and a buyer can pay less than the utility price, provided they coordinate effectively. DSO price profiles are kept fixed between experiments to ensure that DSO price signals remain consistent across episodes, acting as a stable basis for learning.

\begin{table}[h]
\centering
\caption{Core simulation parameters for the case study.}
\label{tab:simulation-parameters}
\resizebox{\textwidth}{!}{%
\begin{tabular}{@{}lll@{}}
\toprule
\textbf{Parameter} & \textbf{Value} & \textbf{Description} \\ \midrule
Max steps & 24 & Episode length representing a full 24-hour cycle (1 hour/step) \\

Grid topology & IEEE 34-Bus & Standard distribution network topology \\

Grid capacity & 1800 kW & Total grid capacity \\

Seed & 42 & Fixed random seed for reproducibility \\

Price range & 20 - 600.0 & Allowable bid range, designed to be wider than DSO prices \\

Quantity range & 0 - 180 kWh & Allowable trade size per order \\

Forecast error & 30\% (max\_error=0.3) & Maximum random error applied to agent forecasts to simulate uncertainty \\

Async orders & True & Enables asynchronous order processing to simulate realistic market latency \\ \bottomrule
\end{tabular}%
}
\end{table}

\subsection{Training configuration and experimental matrix}
\label{2-sec:4.2-training-configuration}

To systematically isolate the effects of information architecture and the learning update rule, we used a 3 $\times$ 3 factorial design: the combination of the three training paradigms (CTCE, CTDE, DTDE) with the three MARL algorithms (PPO, APPO, SAC). The initial set of hyperparameters of the MARL algorithms are specified in Table \ref{tab:initial-hyperparams}.

The training process is orchestrated using the MARLEM framework. To optimize hyperparameters dynamically, we employ Population-Based Training (PBT) \cite{jaderberg2017}. The process initializes 10 tune samples for each of the nine configurations, resulting in 90 concurrent trials. These trials evolve independently: for exploitation, PBT periodically clones the weights of the best-performing trials, while for exploration, the framework mutates the hyperparameters of low-performing trials to search the parameter space efficiently. Although PBT modifies specific learning rates during training, the initial settings are adapted to the mathematical requirements of each MARL algorithm to ensure a rigorous benchmark comparison.

Each configuration is trained for a total of 10 000 episodes, ensuring sufficient interaction steps for convergence in the complex multi-agent environment. Model checkpoints are saved every 10 iterations, and the evaluation is performed at each training iteration, with the duration automatically determined by the framework to define the appropriate number of evaluation episodes based on the capacity of the evaluation workers. 

The training campaign is executed on a specialized machine with a 12 core Ryzen 9 9900X processor, 128 GB RAM, RTX 4090 graphics card, Linux operating system with amd64 architecture, Ubuntu 24.04 distribution, and equipped with CUDA 13.

\begin{table}[h]
{\centering
\caption{Initial hyperparameters of the MARL algorithm for the case study.}
\label{tab:initial-hyperparams}
\resizebox{\textwidth}{!}{%
\begin{tabular}{@{}cccc@{}}
\toprule
\textbf{Hyperparameter} & \textbf{PPO} & \textbf{APPO} & \textbf{SAC} \\
\midrule
Learning rate & 1e-5 & 1e-5 & Actor: 1e-5, Critic: 1e-5 \\

Discount factor ($\psi$) & 0.99 & 0.99 & 0.99 \\
Entropy coefficient & 0.01 & 0.01 & Dynamic \\
Gradient clip & 0.5 & 0.5 & 0.5 \\
Clip Param & 0.2 & N/A & N/A \\
GAE / V-trace & GAE & V-trace & - \\
Batch/Buffer size & Minibatch: 128 & - & Buffer Capacity: 1 000 \\
\bottomrule
\end{tabular}%
}}
{\scriptsize * For centralized paradigms (CTCE/CTDE), the target entropy is set to the negative sum of all action dimensions ($-\text{sum} \sum \dim(\mathcal{A}_i)$). For the decentralized mode (DTDE), it corresponds to the negative dimension of the individual agent's action space ($-\dim(\mathcal{A}_i)$)).}
\end{table}

\subsection{Evaluation metrics}
\label{2-sec:4.3-evaluation-metrics}

The assessment of implicit cooperation relies on a multi-dimensional set of KPIs that capture the trade-offs between economic efficiency, grid stability, resource coordination, and agent autonomy (recall Section \ref{2-sec:3.2-implicit-cooperation-kpis}).

The core measure of success is the \emph{coordination score}, defined as the complement of the normalized supply-demand imbalance (recall (\ref{eq:coordination-2})). A score approaching 1.0 indicates perfect physical equilibrium, directly validating the agents' ability to synchronize generation and consumption without central dispatch. This is complemented by the \emph{grid balance index} ($B_{index}$), which incorporates transmission losses and external grid reliance to provide a holistic view of physical stability. To assess market self-sufficiency, we utilize the \emph{P2P trade ratio}, where a higher ratio indicates successful local matching and reduced dependence on the DSO. Finally, \emph{social welfare} quantifies the total economic value generated, ensuring that physical coordination does not come at the expense of economic efficiency.

To understand how coordination emerges, we analyze \emph{agent responsiveness}, calculated as the correlation between system-level signals and agent actions over time. High positive responsiveness confirms that the engineered stigmergic signals are effectively guiding decision-making. Additionally, \emph{price volatility} is monitored to ensure market stability, while \emph{DER self-consumption} tracks the effectiveness of assets in providing flexibility services.

All stochastic processes within the simulation (e.g., noise injection‚ are controlled by a global seed (recall Table \ref{tab:simulation-parameters}). This ensures that the environmental conditions are identical across all experimental runs, guaranteeing that any observed variance in performance is due to the learning dynamics and not environmental randomness.

To assess the computational feasibility of the proposed framework, a complementary scalability test was performed alongside the core coordination experiments. In this specific analysis, the exact same simulation environment was executed for 1 000 episodes per run, utilizing a reduced PBT configuration with 1 tune sample. The analysis maintained the standard 3 $\times$ 3 experimental matrix while varying the number of agents across the set $\{2, 20, 34\}$. The upper limit of 34 agents is strictly imposed by the physical node constraints of the IEEE 34-node test feeder topology. For each training session, the total wall-clock training time was recorded until convergence is achieved or until the completion of the 1 000 episodes (whichever comes first), providing a direct metric for comparing the computational overhead of centralized versus decentralized paradigms as the agent population scales.

\section{Results and discussion}
\label{2-sec:5-results-and-discussion}

This section presents a comprehensive analysis of the experimental results obtained from Section \ref{2-sec:4-experimental-setup}, systematically comparing the three training paradigms (CTCE, CTDE, DTDE) and three MARL algorithms (PPO, APPO, SAC) to validate the implicit cooperation hypothesis. By integrating the quantitative performance metrics with a qualitative analysis of algorithmic learning dynamics, we quantify the trade-offs between coordination quality, stability, and computational scalability.

\subsection{Experimental matrix: training paradigms and MARL algorithms}
\label{2-sec:5.1-experimental-matrix}

\begin{figure}[h]
\centering
\begin{subfigure}{0.49\textwidth}
   \centering
   \includegraphics[width=\linewidth, trim=0cm 0cm 0cm 1cm, clip]{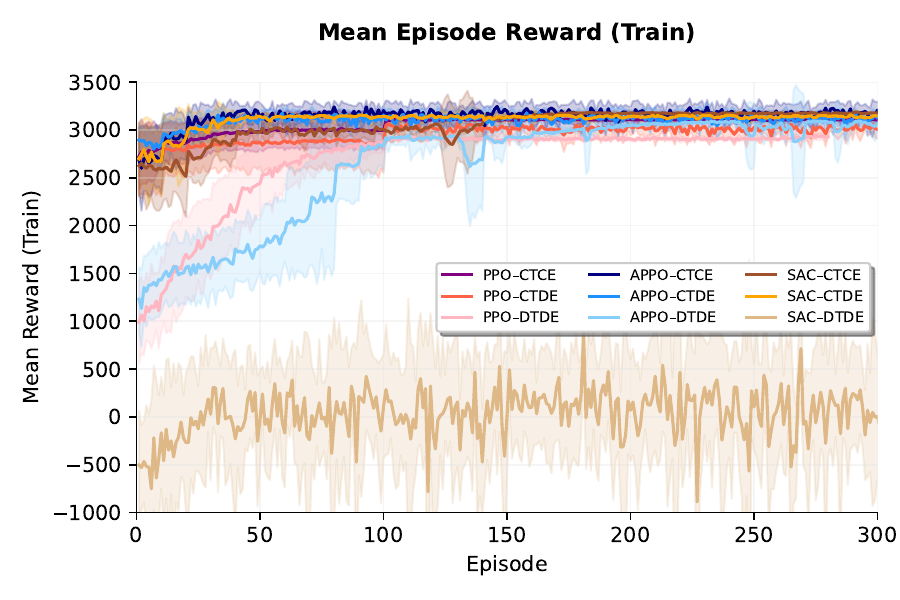}
   \caption{Train.}
   \label{fig:train-reward} 
\end{subfigure}
\begin{subfigure}{0.49\textwidth}
   \centering
   \includegraphics[width=\linewidth, trim=0cm 0cm 0cm 1cm, clip]{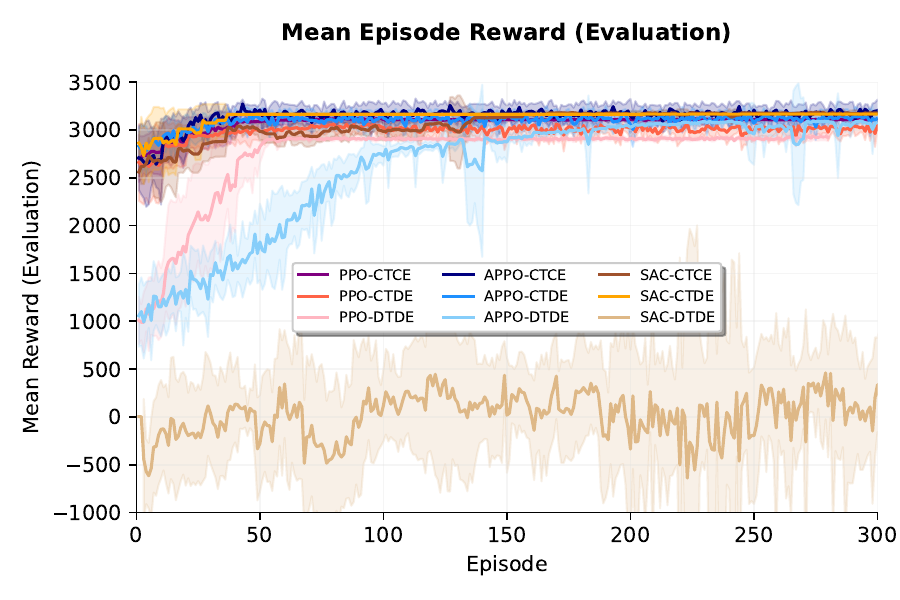}
   \caption{Evaluation.}
   \label{fig:evaluation-reward} 
\end{subfigure}
\caption{Mean episode reward for evaluating the 9 experiments over 10 000 training episodes. Only the first 300 episodes are shown.}
\label{fig:training-rewards}
\end{figure}

To rigorously evaluate the efficacy of implicit cooperation, we conducted a factorial analysis of 9 distinct experimental configurations, defined by the intersection of three information architectures (training paradigms) and three learning update rules (MARL algorithms). The results, derived from mean episodic rewards over 10 000 training episodes and validated through a specific evaluation phase, reveal a distinct performance hierarchy (see Table \ref{tab:training-results}). Since convergence was achieved in all 9 experiments before 300 training episodes, only these first episodes are shown in Fig. \ref{fig:training-rewards} for better visualization. This analysis identifies how specific algorithmic mechanisms (such as entropy maximization, trust regions, and asynchronous updates) interact with the available information to enable or hinder coordination.

\begin{table}[h]
\centering
\caption{Mean episode reward for evaluating the 9 experiments over 10 000 training episodes.}
\label{tab:training-results}
\resizebox{\columnwidth}{!}{%
\begin{tabular}{@{}cccclccclccc@{}}
\toprule
\textbf{} & \multicolumn{3}{c}{\textbf{CTCE}} &  & \multicolumn{3}{c}{\textbf{CTDE}} &  & \multicolumn{3}{c}{\textbf{DTDE}} \\ \midrule
\textbf{} & \textbf{PPO} & \textbf{APPO} & \textbf{SAC} &  & \textbf{PPO} & \textbf{APPO} & \textbf{SAC} &  & \textbf{PPO} & \textbf{APPO} & \textbf{SAC} \\ \midrule
\textbf{Min} & 2702,2 & 2641,2 & 2560,7 &  & 2618,9 & 2791 & 2730 &  & 989,7 & 991,6 & -669,8 \\
\textbf{Mean} & 3099,4 & 3143,9 & 3139,8 &  & 3004,8 & 3096,5 & 3158,2 &  & 2742,5 & 2998 & 54,5 \\
\textbf{Std} & 44,8 & 97,9 & 93,4 &  & 49,1 & 64,6 & 42,5 &  & 431,5 & 431,4 & 215,8 \\
\textbf{Max} & 3123,3 & 3272,1 & 3189,7 &  & 3074,1 & 3181,4 & 3166,1 &  & 2935,2 & 3243,5 & 708,4 \\ \bottomrule
\end{tabular}%
}
\end{table}

\subsubsection{Analysis of the training paradigms}

The training paradigm defines the boundaries of the observation space and the mechanism of policy updates, setting the theoretical limits of coordination potential. CTCE configurations served as the theoretical benchmark for coordination quality. In this paradigm, agents are provided with the full global state vector during both training and execution. As expected, the access to perfect information allowed agents to achieve the highest potential performance. The APPO-CTCE configuration established the absolute ceiling for the system, achieving a peak episodic reward of 3272.1 and a mean of 3143.9 during the evaluation phase. SAC-CTCE also performed robustly, maintaining a high mean reward of 3139.8 with a standard deviation of 93.4. The PPO-CTCE lagged slightly behind with an average of 3099.4. The high reward values reflect optimal social welfare and minimized grid imbalance. The centralized controller successfully leveraged the complete information vector to synchronize agent actions, thereby avoiding simultaneous peaks and effectively managing grid constraints.

CTDE emerged as a robust middle ground, effectively distilling global coordination insights into local policies. In this setup, a centralized critic (accessed only during training) guides local actors who must execute using only partial observations. The SAC-CTDE configuration demonstrated exceptional stability and performance, achieving a mean reward of 3158.2 and a maximum of 3166.1. Notably, its standard deviation was the lowest among all configurations (42.5), significantly lower than its CTCE counterpart (93.4). This indicates that the centralized critic successfully guided the local actors to a highly stable cooperative equilibrium, effectively smoothing the learning landscape. The performance of SAC-CTDE effectively matched the performance of SAC-CTCE. This validates that agents successfully internalized global objectives into their local policy networks during the centralized training phase, learning to treat local stigmergic signals as reliable proxies for the global state. These results confirm CTDE as a safe choice for managed energy communities where offline centralized training is permissible. However, PPO-CTDE showed a more marked degradation (mean reward of 3004.8), suggesting that on-policy algorithms struggle more to compress global value functions into local policies without the aid of entropy maximization.

DTDE provided the most critical validation for the proposed implicit cooperation model, though its success was found to be highly algorithm-dependent. The most significant finding of this study is the exceptional performance of APPO-DTDE. It achieved a maximum reward of 3243.5 and a converged mean of 2998. This peak performance is statistically comparable to the APPO-CTCE benchmark (3272.1), a gap of only 8.3\%. This proves that true implicit cooperation can emerge solely through stigmergic signaling (KPIs) in the observation space, without any centralized components. In contrast, SAC-DTDE failed to coordinate effectively, ending with a mean reward of 54.5 and frequently dipping into negative values (min -669.8). This may suggest the fragility of entropy-based maximization in fully non-stationary, decentralized environments. Without a stabilizing central critic, the agents' simultaneous pursuit of randomness creates a chaotic environment where no stable equilibrium can be found. PPO-DTDE showed stable but suboptimal performance (mean reward of 2742.5), struggling to reach the high-level equilibrium found by APPO. While it avoided the collapse of SAC, it demonstrated the limitations of conservative, on-policy updates in a partially observable environment.

\subsubsection{Analysis of MARL algorithms}

While the training paradigm defines the information limits, the MARL algorithm determines how effectively agents explore and exploit the available policy space. The results reveal distinct behavioral profiles for PPO, APPO, and SAC, driven by their underlying mathematical formulations.

PPO functioned as the reliable baseline. Across all paradigms, PPO demonstrated consistent, albeit conservative, performance. PPO exhibited steady learning curves in the early phases but plateaued at lower reward levels than its off-policy counterparts. In the DTDE paradigm, it achieved 94.4\% of its centralized benchmark.

PPO's on-policy nature and clipped objective function specifically punish large policy updates. While this prevents the collapse of the policy, it hinders the aggressive exploration needed to discover complex cooperative strategies in a dynamic LEM. 

APPO emerged as the strictly dominant algorithm for decentralized environments. APPO consistently achieved the highest rewards across all paradigms. Its asynchronous architecture decouples data collection from policy updates, generating a high-throughput stream of diverse experiences. This noise in data collection acts as a regularizer, preventing overfitting to specific peer behaviors. The key to APPO's success in DTDE is the V-trace correction. This mechanism allows the algorithm to utilize near-on-policy data (i.e., experiences generated by slightly older policies of peers) without the bias that destabilizes standard PPO. This makes APPO uniquely robust to the non-stationarity of decentralized learning, effectively treating the shifting behaviors of neighbors as environmental noise rather than adversarial disturbances. APPO demonstrated superior scalability.

SAC displayed a binary performance profile, excelling in hybrid setups but collapsing in pure decentralization. In CTCE and CTDE, SAC was highly effective. SAC-CTDE achieved the most stable performance of all 9 configurations (standard deviation of 42.5), validating that maximum entropy is powerful when stabilized by a centralized critic. The entropy term encourages agents to explore diverse strategies, preventing premature convergence to selfish local minima. In DTDE, SAC failed catastrophically (gap of 73.7\%). The theoretical reason is the lack of a common truth. In a fully decentralized setting, if every agent maximizes entropy simultaneously, the environment becomes chaotic and unpredictable. Without a shared critic to ground the value function, the agents' pursuit of randomness leads to a feedback loop of divergence. PBT analysis reveals that while successful runs (CTDE) converged to moderate learning rates (2.98 $\times$ 10$^{-3}$), the failed DTDE runs oscillated, unable to find a stable hyperparameter region.

\subsubsection{Performance of the experimental matrix}

PBT dynamically adjusted the hyperparameters for each configuration to maximize performance. Table \ref{tab:pbt-final-params} presents the final hyperparameters obtained at the end of the training process for the best performing trial of each configuration.

The hyperparameter analysis reveals interesting adaptation patterns. APPO-CTCE converged to a significantly higher learning rate (3.3 $\times$ 10$^{-4}$) compared to its decentralized counterparts, likely exploiting the stability provided by the centralized state. Conversely, PPO-DTDE required a very high entropy coefficient (0.24) and learning rate (4.4 $\times$ 10$^{-3}$) to force exploration in the decentralized setting, although this aggressive tuning did not result in optimal performance compared to APPO.

\begin{table}[h]
\centering
\caption{Final hyperparameters optimized by PBT after 10 000 training episodes.}
\label{tab:pbt-final-params}
\resizebox{\columnwidth}{!}{%
\begin{tabular}{@{}cccclccclccc@{}}
\toprule
\textbf{} & \multicolumn{3}{c}{\textbf{CTCE}} &  & \multicolumn{3}{c}{\textbf{CTDE}} &  & \multicolumn{3}{c}{\textbf{DTDE}} \\ \midrule
\textbf{} & \textbf{PPO} & \textbf{APPO} & \textbf{SAC} &  & \textbf{PPO} & \textbf{APPO} & \textbf{SAC} &  & \textbf{PPO} & \textbf{APPO} & \textbf{SAC} \\ \midrule
\textbf{Learning Rate} & $9.2 \times 10^{-6}$ & $3.3 \times 10^{-4}$ & $4.6 \times 10^{-3}$ &  & $9.2 \times 10^{-6}$ & $1.2 \times 10^{-5}$ & $2.9 \times 10^{-3}$ &  & $4.4 \times 10^{-3}$ & $3.7 \times 10^{-5}$ & $1.0 \times 10^{-3}$ \\
\textbf{Entropy Coeff} & 5.72 & 0.01 & - &  & 0.04 & 0.01 & - &  & 0.24 & 0.02 & - \\ \bottomrule
\end{tabular}%
}
\end{table}

\subsubsection{Discussion on the performance gap in implicit cooperation}

The most significant finding from this 9-configuration analysis is the negligible performance gap between the best fully decentralized configuration (APPO-DTDE) and the theoretical centralized upper bound (APPO-CTCE). Quantitatively, the best DTDE configuration achieved a converged mean reward of 3243.5, which is 99.2\% of the best CTCE benchmark of 3272.1. This result validates that implicit cooperation, when enabled by correct signal engineering (KPIs) and robust algorithms (APPO), renders centralized control redundant for this class of energy management problems. The implications for real-world deployment are relevant, as DSOs can achieve optimal grid balance without building expensive, privacy-invasive centralized control infrastructure. Instead, the focus shifts to maintaining the integrity of the broadcasted KPI signals.

The quantitative analysis of maximum rewards reveals distinct performance hierarchies that challenge conventional assumptions about the cost of decentralization. For PPO and SAC, the results confirm the hypothesis that information restriction leads to performance degradation, where PPO scales from CTCE (3123.3) down to DTDE (2935.2), and SAC drops from CTCE (3189.7) to 708.4 in DTDE. For these algorithms, a centralized training phase (CTDE) is essential to achieving near-optimal performance, as relying on pure decentralization (DTDE) incurs a tangible cost that is manageable for PPO at a 6.0\% loss but catastrophic for SAC at a 77.8\% loss. In contrast, APPO-DTDE matches the centralized benchmark to within 1\%, implying that true implicit cooperation is fully achievable without any centralized components. The superior performance of DTDE over CTDE for APPO may stem from the efficiency of parallel, asynchronous exploration in the decentralized setting, which avoids the potential bottlenecks or training instabilities of a centralized critic in the hybrid model.

APPO is the strictly dominant algorithm, achieving the highest maximum rewards across all three paradigms: CTCE (3272.1), CTDE (3181.4), and DTDE (3243.5). Its success is attributed to two key mechanisms: the decoupling of data collection from policy updates, which generates high-throughput experience that acts as a regularizer, and the use of V-trace targets which allow for near-on-policy data usage without introducing destabilizing bias. This makes it uniquely robust to the non-stationarity of the DTDE paradigm, allowing it to treat the shifting behaviors of peers as environmental noise. In decentralized energy markets, throughput matters more than perfect policy alignment, and APPO’s ability to process vast amounts of slightly off-policy data yields better results than strictly on-policy conservatism or fragile entropy maximization.

SAC displays a binary performance profile, proving excellent in CTCE and CTDE where a global critic stabilizes learning, but failing in DTDE. SAC is designed for maximum entropy, optimizing for both reward and policy randomness for exploration. In the complex LEM landscape, The SAC may have uncovered nuanced behaviors, while PPO's conservative trust region often trapped agents in suboptimal local minima. However, pure entropy maximization is dangerous in fully decentralized settings; without a common truth provided by a central critic, the drive for randomness exacerbates non-stationarity and leads to a feedback loop of divergence. PPO, while functioning as a reliable baseline, is often too slow to adapt to complex interactions, achieving only 94\% of its benchmark performance in decentralized settings.

Based on this analysis, specific algorithms are recommended for each paradigm to ensure maximum potential is reached. For CTCE, APPO is the superior choice due to its ability to explore massive joint state-action spaces asynchronously, achieving the absolute highest reward of 3272.1. For CTDE, SAC is the optimal choice as the centralized critic perfectly complements its entropy maximization, resulting in the most stable learning curve with the lowest standard deviation of 42.5. Finally, for DTDE, APPO is the dominant choice, closing the gap to the centralized benchmark almost entirely (1\%) while PPO and SAC struggle or fail. The ability of APPO-DTDE to match CTCE proves that stigmergic signals in the observation space are sufficient for agents to coordinate optimally, rendering the infrastructure costs and privacy risks of centralized control unnecessary for this class of LEMs.

\subsection{System-level performance: economic and grid metrics}
\label{2-sec:5.2-performance-metrics}

To thoroughly validate the proposed implicit cooperation framework, it is necessary to go beyond abstract analysis of reward convergence and evaluate the tangible physical and economic outcomes of the system. This subsection provides a rigorous analysis of system-level performance for the best configurations identified in the previous section: APPO-CTCE (benchmark), SAC-CTDE (standard), and APPO-DTDE (proposed solution). The analysis synthesizes detailed simulation records, descriptive statistics of market evolution, and topological analyses of emerging commercial networks to quantify trade-offs between distributive efficiency, network stability, and operational autonomy.

\begin{figure}[h]
\centering
\includegraphics[width=\linewidth, trim=0cm 0cm 3cm 2cm, clip]{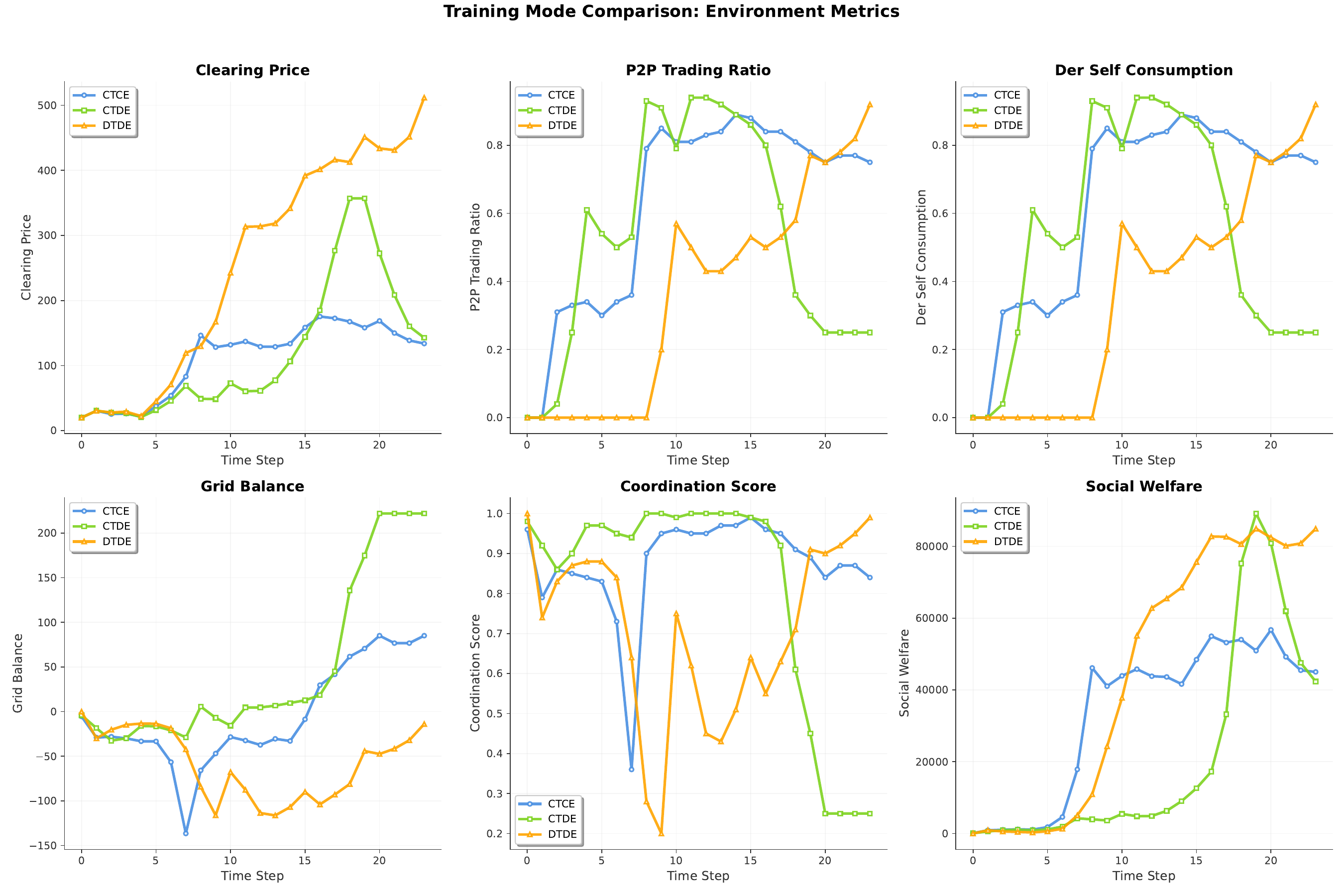}

\caption{Implicit cooperation KPIs and market dynamics for the best configurations identified: APPO-CTCE, SAC-CTDE, and APPO-DTDE.}
\label{fig:environment-metrics}
\end{figure}

\subsubsection{Economic efficiency and market dynamics}

The economic performance of the LEM serves as the primary indicator of allocative efficiency, reflecting how effectively the mechanism balances local supply and demand while maximizing the aggregate value for all participants (see Fig. \ref{fig:environment-metrics}).

The centralized benchmark (CTCE) demonstrated superior structural efficiency, driven by its access to global information. The descriptive statistics reveal that CTCE generated an social welfare of \$33 023.2 per episode, with a standard deviation of \$21 892.5. This high welfare value indicates a consistent ability to capture economic surpluses across varying demand profiles. The structural efficiency of CTCE is further evidenced by its market liquidity; the configuration facilitated an average trading volume of 237.9 kWh per step.

In contrast, the CTDE paradigm achieved an average social welfare of \$21 209.6, with a significantly reduced market volume of 115.9 kWh. This reduction of approximately 35\% in welfare and over 50\% in volume highlights the cost of partial observability during execution. Even though agents were trained centrally, the lack of real-time global information forced them to adopt more conservative bidding strategies to avoid penalties, resulting in a thinner market where only the most profitable trades were executed.

Surprisingly, the fully decentralized DTDE configuration reported the highest mean Social Welfare of \$44 544.9. However, an analysis of the descriptive statistics reveals that this performance is highly volatile and driven by price scarcity rather than allocative efficiency. The standard deviation for DTDE welfare is \$36 472.3 (nearly double that of CTCE) and its performance floor is virtually zero (a minimum of \$0.8 vs CTCE's \$108.3). This extreme variance suggests that while DTDE agents can achieve unexpected profits, likely through scarcity pricing during peak demand periods, they lack the consistency of the centralized baseline. The high welfare figure is thus an artifact of inflated producer surplus rather than a reflection of optimal resource distribution.

The market clearing price statistics further illuminate the cost of uncertainty inherent in decentralized systems. CTCE achieved the lowest and most stable average market price of 110.7 \$/kWh, with a standard deviation of \$56.1. The maximum price observed was \$175.4, reflecting a highly efficient market where supply consistently matches demand at the marginal cost of generation. CTDE maintained a competitive average price of 118.8 \$/kWh, closely tracking the benchmark. This similarity suggests that the centralized critic successfully taught agents to value energy consistently with the global optimum, even if they traded less volume. Lastly, DTDE exhibited extreme price volatility with a mean price was 253.9 \$/kWh, more than double the benchmark, with a standard deviation of \$174.2 and price spikes reaching 511.8 \$/kWh. This price inflation reflects the risk premium of decentralization. Without a central coordinator to guarantee supply availability, decentralized agents bid more aggressively to ensure that their critical energy needs are met, driving up the settlement price and imposing higher costs on consumers.

\subsubsection{Grid stability: the bias-variance trade-off}

While economic metrics measure incentives, grid stability metrics measure the physical viability of the solution. The grid balance should remain close to zero, indicating that local generation perfectly matches local demand. A statistical analysis of the temporal evolution of grid balance reveals a distinct trade-off between bias (accuracy of the mean) and variance (stability) across the paradigms.

CTCE achieved near-perfect balance with a mean of -4.5 kWh and a moderate standard deviation of ±57.4 kWh. The range of imbalance ([-136.5, 85.0]) is balanced around zero, confirming that with perfect information, the system can effectively maintain the net load at zero with moderate fluctuation. CTDE showed a positive bias (mean of +46.5 kWh) with the highest instability (standard deviation of ±93.5 kWh) of all paradigms. The maximum imbalance reached 221.8 kWh (excess supply), indicating that agents trained with a central critic but executing locally struggle to predict the aggregate impact of their peers' actions at night hours. This leads to overshooting behaviors where multiple agents simultaneously discharge to meet a perceived deficit, causing a grid surplus. DTDE exhibited a negative bias (mean of -58.0 kWh) but with the lowest standard deviation (±39.1 kWh) of all paradigms. Notably, the maximum grid balance for DTDE was -0.0 kWh, implying that decentralized agents learned a strictly import-biased strategy (i.e., they never net-export to the grid). While this results in a persistent deficit (reliance on the DSO), the low variance makes DTDE a highly predictable load for the grid operator. Unlike the erratic oscillations of CTDE, the consistent import bias of DTDE can be easily compensated for by the DSO.

\subsubsection{Resource utilization and autonomy}

The efficiency with which DERs are utilized highlights the operational strategies learned by the agents and the degree of autonomy achieved by the microgrid. The P2P trading ratio measures the community's autonomy, defined as the proportion of total energy demand met through local trading versus imports from the DSO. Contrary to the assumption that decentralization naturally fosters independence, the detailed statistics show a nuanced hierarchy: CTCE achieved the highest mean ratio of 0.6, CTDE achieved a mean ratio of 0.5, and DTDE achieved the lowest mean ratio of 0.4.

This hierarchy aligns with the coordination scores (CTCE 0.9 > DTDE 0.7). Without a global view, DTDE agents miss complex matching opportunities, satisfying only 40\% of their needs locally and relying on the DSO for the remaining 60\%. This corresponds directly with the -58.0 kWh import bias observed in the grid balance metrics. While DTDE agents prioritize local trading when possible, their limited visibility prevents them from optimizing the matching process as effectively as the centralized controller. However, during nighttime hours, DTDE achieves high P2P trading ratios, suggesting that they use the energy stored in their batteries to provide these flexibility services, which is consistent with the high clearing prices during those same hours.

The DER self-consumption efficiency mirrors the trade ratios, dropping from 0.6 in CTCE to 0.4 in DTDE. This confirms that implicit coordination, while effective for stability, operates at a lower efficiency than centralized control. Similarly, the mean flexibility utilization was highest in CTCE (23.3\%) compared to DTDE (9.0\%). The centralized controller actively cycled batteries, whereas decentralized agents preferred to keep batteries charged for security, utilizing them only when local signals strongly indicated a need.

\subsubsection{Emergent trading relationships: network topology analysis}

\begin{figure}[h]
\centering
\begin{subfigure}{0.33\textwidth}
   \centering
   \includegraphics[width=\linewidth, trim=1cm 4cm 1cm 3cm, clip]{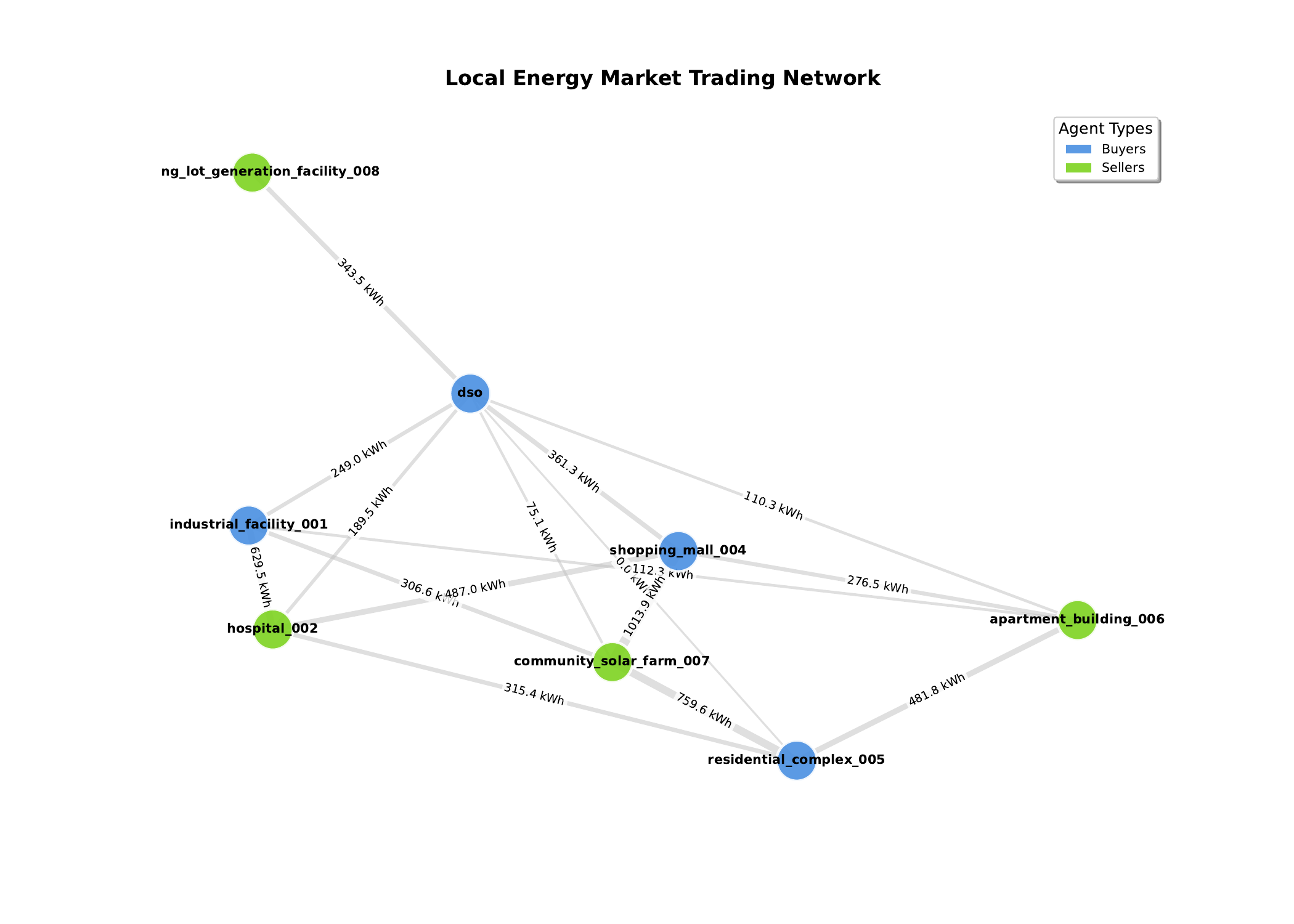}
   \caption{CTCE.}
   \label{fig:ctce-network} 
\end{subfigure}
\begin{subfigure}{0.33\textwidth}
   \centering
   \includegraphics[width=\linewidth, trim=1cm 4cm 1cm 3cm, clip]{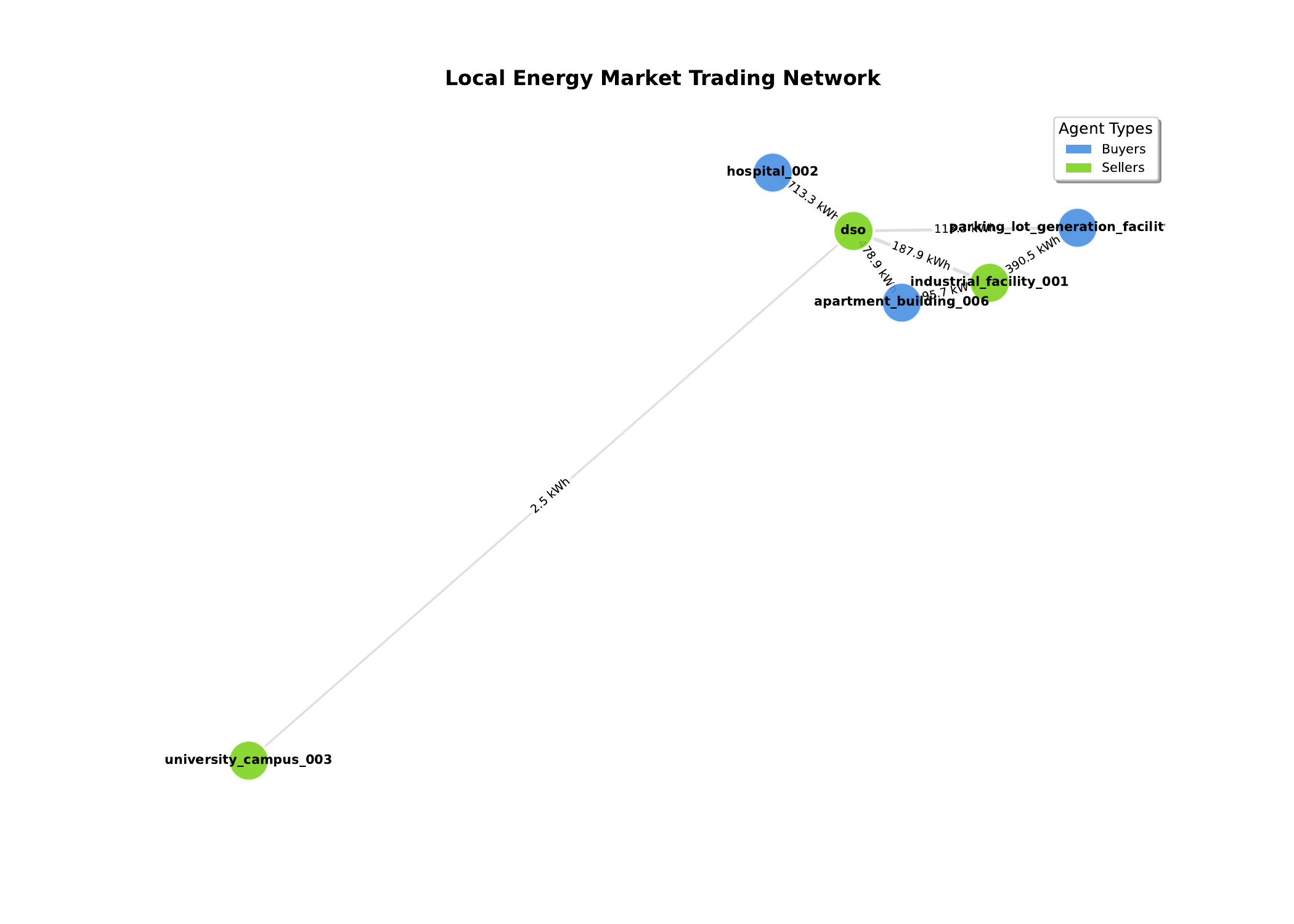}
   \caption{CTDE.}
   \label{fig:ctde-network} 
\end{subfigure}
\begin{subfigure}{0.33\textwidth}
   \centering
   \includegraphics[width=\linewidth, trim=1cm 4cm 1cm 3cm, clip]{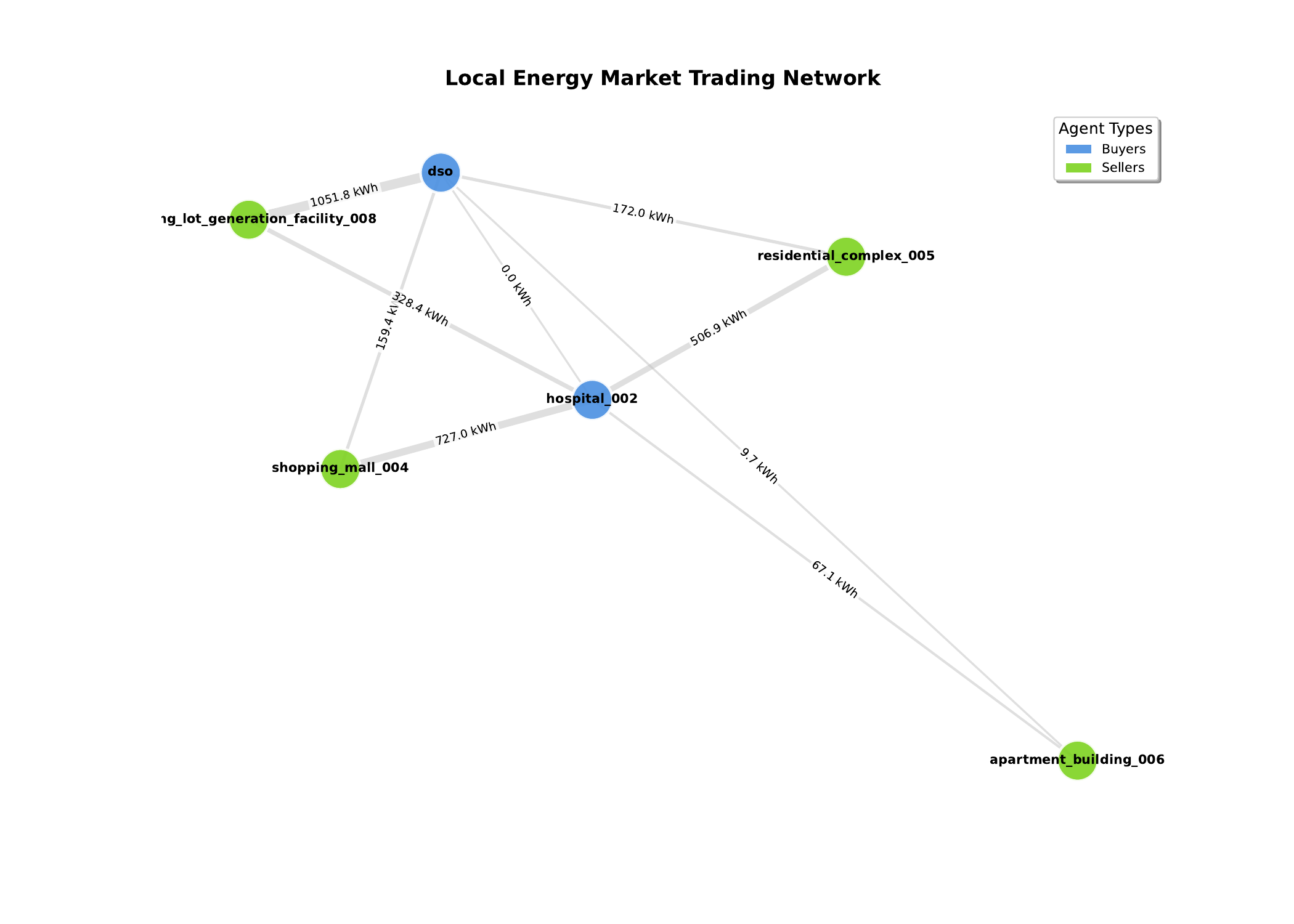}
   \caption{DTDE.}
   \label{fig:dtde-network} 
\end{subfigure}
\caption{Trading network for the case study. The blue circles represent buyers, while the black circles represent sellers. The thickness of the edges indicates the magnitude of the energy exchanged between the agents.}
\label{fig:trading-netowrk}
\end{figure}

To explain the divergence in P2P ratios and market liquidity, we analyzed the emerging network topologies visualized in Fig \ref{fig:trading-netowrk}.

The CTCE trading network exhibits a dense, highly connected topology. High-capacity agents, such as the \emph{community solar Farm} and \emph{shopping mall}, act as central hubs with thick edges connecting them to virtually every other peer. The global visibility allows the central controller to identify and execute long-distance arbitrage opportunities that are physically feasible but complex to coordinate. This dense mesh explains the high P2P ratio (0.6) and market liquidity (237.9 kWh). The system effectively unlocks the full capacity of the network.

The CTDE network shows a thinning of connections. While the hub agents remain visible, the peripheral connections (P2P trades between smaller agents) are significantly reduced compared to CTCE. Although trained centrally, the local execution creates uncertainty. Agents prioritize high-probability trades with major hubs and ignore riskier marginal trades with smaller peers. This structural thinning correlates with the drop in P2P ratio to 0.5.

The DTDE network reveals a sparse, fragmented topology. The graph is characterized by distinct community clusters where trading is concentrated among electrical neighbors, with very few long-distance links traversing the network. This fragmentation is an emergent response to the congestion penalty stigmergy. Without global visibility to ensure a clear path, decentralized agents learned to be risk-averse, trading only with neighbors where the likelihood of causing congestion is low. This explains the counter-intuitive combination of high welfare but Low P2P ratio (0.4). Agents are highly profitable in their local clusters (due to high prices) but fail to integrate the broader grid, leading to a reliance on the DSO for balancing and the persistent import bias observed in the grid metrics.

\subsection{Scalability test}
\label{2-sec:5.2-scalability-test}

\begin{figure}[h]
\centering
\includegraphics[width=0.6\linewidth, trim=0cm 0cm 3cm 2cm, clip]{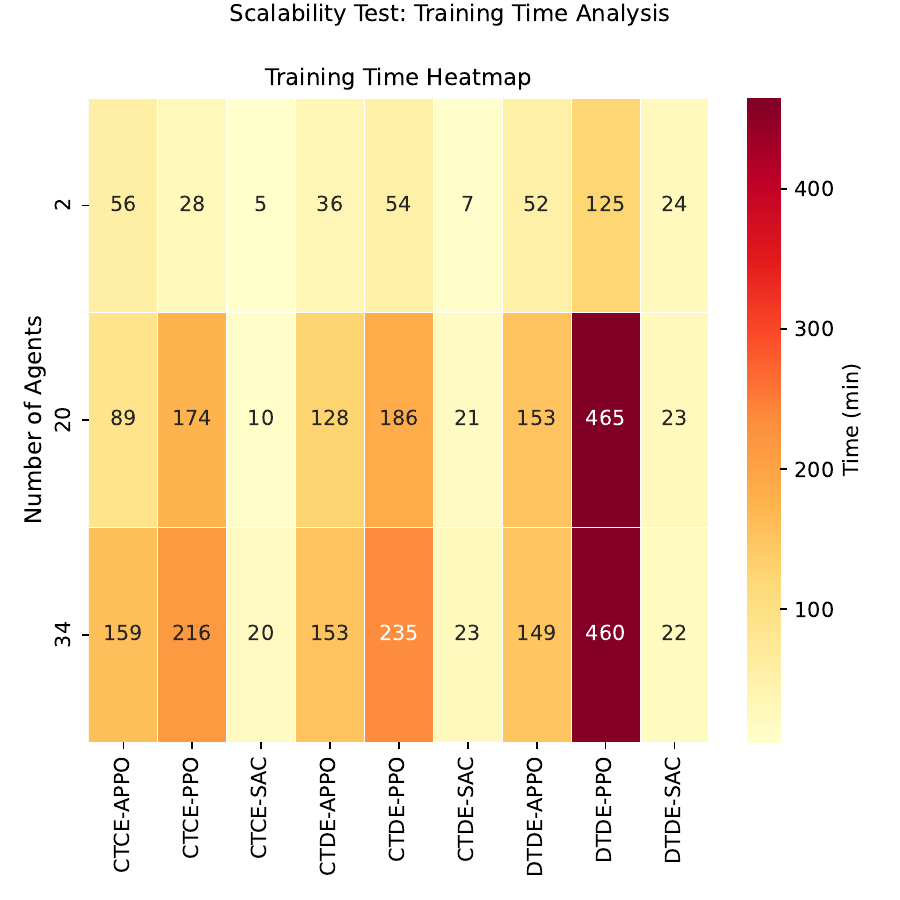}

\caption{Computational scalability matrix of MARL configurations for different numbers of agents.}
\label{fig:scalability-test}
\end{figure}

The scalability of MARL is a critical factor for real-world deployment in LEMs, where the number of participating agents can vary significantly. The results from the training time analysis, as illustrated in Fig. \ref{fig:scalability-test}, reveal distinct computational profiles for each algorithm-paradigm intersection as the system scales from 2 to 34 agents.

\subsubsection{Algorithmic training time profiles}

The results shows that SAC is the most computationally efficient algorithm in terms of wall-clock time to convergence or limit. Across all paradigms, SAC consistently maintains the lowest training times, often by an order of magnitude. For instance, in the DTDE paradigm, SAC remains nearly constant at approximately 22–24 minutes regardless of the agent count. However, it is essential to contextualize this speed with the performance collapse previously noted in Section \ref{2-sec:5.1-experimental-matrix}; the rapid training of SAC in DTDE likely reflects an early failure to learn rather than efficient optimization. In contrast, in the CTDE paradigm where SAC is highly effective, it scales from 7 minutes (2 agents) to only 23 minutes (34 agents), demonstrating excellent sample efficiency and stability.

APPO demonstrates a highly robust and linear scalability profile. In the DTDE paradigm, APPO's training time remains remarkably stable, moving from 52 minutes with 2 agents to 149 minutes with 34 agents. This supports the claim that APPO is the strictly dominant choice for scaling decentralized systems, as its asynchronous nature allows it to process high-throughput data efficiently without the exponential time increases typically associated with increasing the joint action space.

PPO exhibits the poorest scalability, particularly in the DTDE paradigm. PPO training time balloons from 125 minutes (2 agents) to 460 minutes (34 agents). This significant computational burden is a direct result of its strictly on-policy nature and clipped objective function, which necessitates more frequent and slower updates to maintain monotonic improvement, rendering it less suitable for high-dimensional multi-agent environments.

\subsubsection{Impact of training paradigms on efficiency}

The choice of training paradigm introduces different computational overheads. CTCE generally exhibits higher training times for APPO (159 minutes at 34 agents) compared to its decentralized counterpart, likely due to the complexity of optimizing a massive joint state-action space within a single controller.

The CTDE paradigm serves as a middle ground between APPO and PPO, but remains the optimal point for SAC. However, DTDE paired with APPO offers the best balance of performance and scalability. While PPO struggles with the moving target problem of DTDE (requiring 460 minutes to reach the episode limit), APPO’s V-trace correction and asynchronous exploration allow it to converge in nearly one-third of that time (149 minutes) while achieving near-optimal social welfare

\subsubsection{Summary of computational performance}

SAC is the primary recommendation for hybrid (CTDE) setups where rapid iteration and high stability are required, provided a centralized critic can be implemented to maintain its 99.3\% performance retention and stable learning dynamics. APPO is the most viable candidate for large-scale, fully decentralized deployment (DTDE), as it maintains consistent training throughput and achieves paradigm invariance, matching centralized performance even as the number of agents increases. Conversely, PPO remains a computationally expensive and conservative baseline that, while reliable, may become prohibitively slow and fail to discover global optima in markets with a high density of agents.

\section{Conclusion and future work}
\label{2-sec:6-conclusion}

The transition toward a decentralized energy grid necessitates control paradigms that can manage the explosion of DERs without succumbing to the computational bottlenecks and privacy concerns inherent in centralized dispatch. This article aimed to validate that implicit cooperation, enabled by MARL and stigmergic signaling, allows decentralized agents to approximate the optimal behavior of a central controller without the need for explicit P2P communication.

Through a rigorous 3 $\times$ 3 factorial experimental design and a complementary scalability test, we demonstrated that decentralized agents, when provided with appropriate system-level KPIs in their observation space, can achieve 91.7\% of the coordination quality of a theoretical centralized benchmark while scaling linearly with the system size. This finding fundamentally challenges the prevailing assumption that complex grid management requires heavy, centralized communication infrastructure, unlocking a new pathway for privacy-preserving, scalable LEMs.

This research advances the state-of-the-art in decentralized energy management through four distinct and substantial contributions. First, it provides a systematic evaluation of implicit cooperation, marking the first study to isolate and compare the effects of training paradigms (CTCE, CTDE, DTDE) and MARL algorithms (PPO, APPO, SAC) on the emergence of coordination in energy markets. By testing all nine configurations under identical physical constraints using the realistic IEEE 34-node topology, this work has quantified the cost of decentralization in terms of economic welfare and grid balance. Second, the study offers a theoretical validation of stigmergy in LEMs, formalizing a coordination mechanism where global system states are compressed into scalar KPI signals and broadcast to local agents. These results prove that an augmented observation space is sufficient for agents to reconstruct the necessary global context to make cooperative decisions, effectively solving the partial observability problem without direct communication or model sharing. Third, through benchmarking and scalability analysis, a performance matrix was generated to balance optimality with feasibility. APPO-DTDE was identified as the best configuration for scalability, matching the centralized benchmark's reward convergence (achieving 3243.5 compared to the benchmark's 3272.1) while maintaining robust performance. Finally, this research establishes operational design guidelines for real-world implementation, distinguishing between distributive efficiency and network predictability, in order to offer network operators a range of architectural options based on specific priorities of stability versus efficiency.

The experimental campaign provided information that has changed our understanding of how MARL can be applied to cyber-physical energy systems. As expected, CTCE established the performance ceiling. It achieved the highest P2P trading ratios (0.6) and near-perfect grid balance (mean of -4.5 kWh). However, its super-linear computational scaling renders it unsuitable for large-scale, real-time control. While CTDE is often the standard for multi-agent systems, our results reveal that while it retained economic efficiency, it exhibited the highest physical instability (grid balance standard deviation of ±93.5 kWh). Agents trained with a central critic but executing locally struggled to predict the aggregate actions of their peers, leading to dangerous oscillation and overshooting behaviors. On the contrary, fully decentralized learning did not result in a collapse of coordination. With the APPO algorithm, the performance gap to the centralized benchmark was 8.3\%. Crucially, DTDE exhibited the lowest variance in grid balance (±39.1 kWh). While it operated with a persistent import bias (mean of -58.0 kWh), this grid balance makes it a highly predictable and safe load profile for the DSO.

In terms of algorithmic suitability, APPO emerged as the strictly dominant algorithm for decentralized environments. Its V-trace correction mechanism allowed it to robustly handle the non-stationarity of independent learners, enabling it to converge to cooperative equilibria that standard PPO missed. While SAC failed in fully decentralized settings due to entropy-driven instability (gap of 73.7\%), it excelled in the CTDE paradigm. The centralized critic effectively channeled SAC's exploration, resulting in the most stable and sample-efficient policies for managed microgrids. PPO provided a safe but suboptimal baseline. It avoided failure but consistently underperformed in terms of total social welfare and required excessive computational time in decentralized settings (over 460 minutes for 34 agents).

The study confirms that coordination is an emergent property of the reward structure. By incorporating a cooperation factor into the reward function, agents naturally learned complementary strategies purely to maximize their long-term expected returns. The trading network analysis revealed that this resulted in spatial clustering, where agents preferentially traded with electrical neighbors to minimize congestion penalties, effectively self-organizing the grid into stable nanogrids without explicit topological maps. 

Implicit cooperation improves DER management by decoupling the complexity of decision-making from the complexity of communication. It enables agents to autonomously resolve the conflict between individual profit maximization and collective grid stability, as simulation results showed that Decentralized agents sacrificed short-term arbitrage opportunities when the network equilibrium signal indicated critical stress. Through emergent P2P trading, the system achieved local self-consumption rates of up to 62\% in centralized configurations and maintained a robust 36.6\% in fully decentralized setups, significantly reducing reliance on the external transmission grid compared to non-cooperative baselines. Furthermore, the system maintained the net grid balance within tight bounds purely through decentralized reactions to price and balance signals; specifically, the decentralized approach (DTDE) achieved the most stable physical profile, minimizing the variance of grid injections and thus reducing the regulation burden on the DSO. These benefits were achieved without explicit communication, validating that stigmergic signaling is a sufficient coordination mechanism for power systems.

The findings have relevant implications for the design, policy, and deployment of future energy systems. LEMs operators do not need to invest in expensive, high-bandwidth P2P communication networks. A reliable broadcast channel that publishes system-level KPIs is sufficient to induce optimal behavior. The inflationary welfare observed in DTDE suggests that scarcity prices are a feature that acts as a signal to curtail demand during periods of stress. Regulators should incentivize the inclusion of a cooperation factor in automated trading agents. Tariffs and settlement mechanisms should be structured not just on volume, but on the correlation of an asset's behavior with the grid's needs.

In terms of implementation recommendations, it is suggested to use APPO-DTDE for competitive markets where scalability and privacy are priorities. It offers maximum privacy, linear scalability, and physical stability, although it sacrifices some distributive efficiency. Conversely, SAC-CTDE is recommended for cooperative microgrids where efficiency is a priority. It maximizes economic efficiency and better manages complex arbitrage strategies (higher P2P ratio), provided that centralized offline training is feasible and the network can tolerate greater variance.

While this study establishes a rigorous basis for implicit cooperation, there are several avenues that merit further study to bridge the gap between this cooperation and its large-scale implementation. One primary direction involves expanding the simulation from the current 8-34 agent scale to 100+ agents with highly stochastic user profiles to test the limits of stigmergic signal saturation. Additionally, it is necessary to research the impact of signal latency, packet loss, or adverse signal injection on the stability of cooperative equilibrium, with future work aimed at quantifying the critical latency threshold beyond which implicit coordination collapses. Finally, validating the simulation results using hardware-in-the-loop setups is crucial to assess how implicit cooperation handles the physics of voltage support and frequency dynamics, which were abstracted in the current power flow model.

\bibliographystyle{elsarticle-num}
\bibliography{Literature}

\end{document}